\documentclass[12pt]{article}
\usepackage{amsmath}
\usepackage{amssymb}
\usepackage{titlesec}
\titlespacing*{\section}{0pt}{6pt}{6pt}
\titleformat{\subsubsection}[runin]
  {\normalfont\normalsize\bfseries}
  {\thesubsubsection}{1em}{}

\title{\vspace{-4em} Measuring Public Institutions and their Agents: Disparities and Agreement, with Application to Asylum Decisions in the U.S. Immigration Courts\thanks{We thank audiences at the 2026 Joint Statistical Meetings, the University of Chicago Department of Statistics, and the Virginia Chapter of the American Statistical Association. We also thank Jonas Mikhael, Chris Harshaw, Eric Auerbach, David Kallick at the Immigration Research Initiative, and  Adriel Orozco and Aaron Reichlin-Melnick at the American Immigration Council.}}

\author{
\\
Jonathan Auerbach\thanks{Department of Statistics, George Mason University, Fairfax, VA USA, jauerba@gmu.edu} \\
\and
\\
Michael E. Sobel\thanks{Department of Statistics, Columbia University, New York, NY USA, mes105@columbia.edu} 
}

\date{}

\usepackage[letterpaper,margin=1in]{geometry}
\usepackage{setspace}

\usepackage{enumitem}
\setlist[itemize]{itemsep=1em}

\usepackage[
    citestyle=authoryear-comp,
    style=authoryear,
    sorting=nyt,
    labelnumber=true,  
    uniquename=init,
    natbib=true,
    backend=biber,
    giveninits=true,
    maxnames=999,
    maxcitenames=3
]{biblatex}

\DeclareNameAlias{author}{family-given}
\renewbibmacro*{in:}{} 

\usepackage{abstract}

\usepackage{float}
\usepackage{booktabs}
\usepackage{makecell}

\usepackage{graphicx}
\usepackage{subcaption}
\usepackage{placeins}

\usepackage[normalem]{ulem}

\begin{document}
\maketitle
\vspace{-2em}
\setstretch{1.5}
\begin{abstract} 
\vspace{-1em}
\noindent
Public institutions derive their legitimacy from the professionalism and impartiality of their agents. In order to maintain standards of conduct, the performance of an agent is often compared to the performance of her peers. But public institutions typically assign agents to different tasks, and as a result, na\"ive comparisons may reflect the difficulty of the assignment and not an agent's performance. To address this concern, we develop measures of disparity and agreement that compare an agent's performance with the potential performance of randomly selected peers who could have been assigned the same set of tasks. We then give realistic conditions under which disparity is identified from the marginal distribution of each agent’s actions. We also show how agreement, which is generally unidentified when agents are assigned to different tasks, can be bounded or estimated under sensitivity assumptions about the dependence between potential outcomes. Finally, we use a nonparametric Bayesian model to examine applications for asylum filed between 2007 and 2014 in the widely studied and often maligned U.S. immigration courts. Our results do not support the picture of widespread partiality asserted in previous empirical work. At both the judge and court level, we find limited disparity and substantial potential agreement.
\end{abstract}
\vspace{1em}
\vfill

\setstretch{1.7}
\begin{refsection}

\section{Introduction} 

Public institutions are more likely to be perceived as trustworthy when they hold their agents accountable. To maintain accountability, it is necessary to evaluate agent performance.

If performance criteria are clear and easy to use, as when 
caseworkers from child protective services are evaluated on whether they made the mandatory number of visits to their clients last month, an audit study can determine if this condition was met. Similarly, teachers may be evaluated on whether a given percentage of students passed a standardized exam.

Yet even when criteria are clear, it is often difficult to determine whether they have been met. For example, officers are prohibited from using excessive force: ``Officers may use force only when no reasonably effective, safe, and feasible alternative appears to exist and may use only the level of force that a reasonable officer on the scene would use under the same or similar circumstances'' \citep{dojforce}. 
Similarly, U.S. immigration judges are required to grant asylum to applicants who complied with procedural criteria and demonstrate a well founded fear of persecution based on protected grounds (race, religion, nationality, political opinion or membership in a protected social category). 

But how are we to assess whether an officer's behavior is reasonable, or if an immigration judge has improperly denied admission to asylum seekers? That is the issue we address.

One approach is to use a `gold standard,' measuring whether an agent's performance on a set of model cases is consistent with the behavior expected under ideal conditions. But even though every agent is measured against the same standard, the model cases may not be representative of the situations that at least some agents encounter, as when difficult or easy cases are underrepresented. Moreover, agents may know how to respond to model cases, but behave differently in real life situations.
 
A second approach uses a `community standard.' Here, there are no model cases. Instead, an agent's behavior warrants further scrutiny (for outstanding or poor performance) 
if it deviates from the behavior of peers in actual situations. Although the community standard does not capture systemic departures from ideal behavior, as would occur if all agents discriminate in the same way, it is easy to implement---at least in theory---and widely used.

Yet determining whether an agent behaves like her peers is not always straightforward in practice. Encounters with the public are typically handled by just one agent. For example, one police officer makes a particular traffic stop. An evaluation using a community standard might compare officers according to the rate at which they use force in those stops, but then the challenge is to decide which other officers or stops make for a fair comparison. Simple disparity measures, such as the average difference in outcomes between agents and peers, can be misleading. In addition, such comparisons do not indicate whether agents and peers agree: this requires information on the unobserved joint distribution of potential outcomes.

This paper develops methodology for using community standards to assess the performance of public agents and the institutions they represent, focusing on the common situation in which each instance is handled by one agent. 
Our methodology is also applicable in other settings, e.g., large firms.

We proceed as follows. Measures of disparity, defined as a difference in performance outcomes between 
an agent 
and her peers in comparable situtations, are given in Section 2. While these can be identified from the marginal distributions of outcomes under each agent, agreement between the outcomes is typically of greater interest. 
Therefore, we also define measures of agreement, which depend on the unobserved joint distribution of agents' potential behavior and are only partially identified from the marginal distributions. We tighten the traditional Fr\'echet-Hoeffding (herafter FH) bounds under realistic assumptions about the evidence different agents see and how that evidence would be handled by another agent viewing the same case. We also outline additional conditions for estimation.

Our application to the U.S. immigration courts is presented in Section 3. We begin by reviewing several influential studies that espoused the view, now widely held, that variation between courts and between judges within courts in the rate at which asylum seekers were granted reflect a fundamental lack of institutional consistency. We show that this work confuses disparities with agreement and fails to properly account for the fact that different judges adjudicate different types of cases. We then apply our methodology to data on asylum applications filed from May 2007 through September 30, 2014, using a non-parametric Bayesian model to estimate disparities between federal immigration judges and to bound the agreement between them. Our analyses suggest the conclusions in previous work are exaggerated and in need of tempering. Section 4 concludes.

\section{Methodology: Estimands, Identification, and Estimation}
We define estimands for disparity and agreement that compare the performance of agents 
with their peers. We consider the identification or partial identification of these estimands for the setting in which each case is handled by one agent only. We then discuss estimation.

We use the following notation. Let $i \in \{1,...,N\} = \cal{I}$ index the set of cases, $j \in \{1,...,J\} = \cal{J}$ the agents who `treat' cases in $\cal I$, ${\cal I}_{j}$ the set of cases assigned to agent $j$, and ${\cal J}_{i} \subseteq {\cal J}$, with 
two or more members, the set of agents `eligible' to treat case $i \in {\cal I}$. 
Each case is assigned to one agent: ${\cal I}_{j} \bigcap {\cal I}_{j^{\prime}} = \emptyset$, $j \neq j^{\prime}$. Let $Z_{i}$, taking values $z_{i} \in {\cal J}_{i}$, denote the agent assigned to case $i$. Also associated to each case are stochastic potential outcomes $Y_{i}(j)$, $j \in {\cal J}_{i}$, an observed outcome $Y_{i} = \sum_{j \in {\cal J}_{i}}1\{Z_{i} = j\}Y_{i}(j)$ that is continuous or binary, and a vector of recorded covariates ${\bf V}_{i}$ that does not depend on the agent assigned case $i$.

We make the following assumptions, already reflected in the notation above:
\begin{itemize}
\item[{\bf A1}.] No Interference:  For all $i \in {\cal I}$ and for any assignment vectors $(z_{1},...,z_{N})$ and $(z^{\prime}_{1},...,z^{\prime}_{N})$
with $z_{i} = z_{i}^{\prime}$, $Y_{i}(z_{1},...,z_{N}) = Y_{i}(z^{\prime}_{1},...,z^{\prime}_{N}) \equiv Y_{i}(z_{i})$.
\item[{\bf A2}.] Consistency: For all $i \in \cal{P}$ and $j \in \cal{J}$, $Z_{i} = j$ implies $Y_{i} = Y_{i}(j)$.

Under assumption \bf{A2}, $\mathbb{E}[Y_{i} \mid Z_{i} = j, {\bf V}_{i} = {\bf v}] =  
\mathbb{E}[Y_{i}(j) \mid Z_{i} = j, {\bf V}_{i} = {\bf v}]$.
\end{itemize}

\subsection{Disparity}
\subsubsection*{Measures}
Many measures have been used to quantify the disparity or inequality between different sub-groups of a population, e.g., men and women \citep{duan2008disparities, cook2012measuring}.
Our goal is different. We aim to construct `local' measures for comparing the performance of an agent $j$ with her peers and  `global' measures of institutional performance that take into account the behavior of all agents in $\cal J$.

Previous researchers (e.g., \cite{GAO2016,ramji2007}) have used
\begin{equation}
\text{D}(j,j^{\prime}) =  N_{j}^{-1}\sum_{i \in {\cal I}_{j}}Y_{i} -  N_{j^{\prime}}^{-1}\sum_{i \in {\cal I}_{j^{\prime}}}Y_{i},
\label{eq:disparity1}
\end{equation}
the average difference for outcomes treated by $j$ versus outcomes treated by $j^{\prime}$, and 
\begin{equation}
 \text{D} (j,\cdot) =    N_{j}^{-1} \sum_{i \in {\cal I}_{j}} Y_{i} -  N^{-1} \sum_{i \in {\cal I}} Y_{i},
\label{eq:disparity2}
\end{equation}
the average deviation of $j$ from the institutional average, as disparity measures, where $N_{j}$ is the cardinality of ${\cal I}_{j}$.

In a completely randomized experiment with only two agents $j$ and $j^{\prime}$, (\ref{eq:disparity1})  is an unbiased estimator of the average effect of assigning $j$ versus $j^{\prime}$ to the cases in $\cal{I}$.  But often agents are not randomly assigned to cases, and even when they are, they may not be eligible for assignment to the same cases. In such instances, (\ref{eq:disparity1}) may not indicate how $j$ and $j^{\prime}$ differ were they assigned similar sets of cases. This suggests developing measures that compare agent $j$ only with peers who are eligible to be assigned to the cases  $j$ is assigned.

Therefore, we define several local disparity measures that compare agent $j$ to her peers.  
The first is the average (taken over the distribution of cases assigned to agent $j$) effect of assigning $j$ versus an eligible alternate $j^{a}$, randomly selected from ${\cal J}_{i} \setminus j$ for each case $i$:

\begin{equation}
D_{1}(j) = N_{j}^{-1} \sum_{i \in {\cal I}_{j}}\mathbb{E}[Y_{i}(j) - Y_{i}(j^{a})\mid {\bf V}_{i} = {\bf v}_i].
\label{eq:D1}
\end{equation}

The second extends the measure (\ref{eq:D1}) to the set of cases $E_{j}$ that $j$ is eligible to treat,
\begin{equation}
D_{2}(j) = \left (N_{j}^{E} \right )^{-1}\sum_{i \in E_{j}}{\mathbb E}[Y_{i}(j) - Y_{i}(j^{a}) \mid {\bf V}_{i} = {\bf v}_i],
\label{eq:D2}
\end{equation}
where $N_{j}^{E}$ is the cardinality of  $E_{j}$.

Analogously to the difference between the ``effect of treatment on the treated'' and the average treatment effect, $D_{1}(j)$ and $D_{2}(j)$ address somewhat different questions and may differ sharply, as is evident by re-expressing
$D_{2}(j)$ as
\begin{equation} 
D_{2}(j)= \left (N_{j}^{E} \right )^{-1} \Big ( N_{j}D_{1}(j) + \sum_{i \in E_{j} - {\cal I}_{j}} \!\!\!\! \mathbb{E} [Y_{i}(j) - Y_{i}(j^{a}) \mid {\bf V}_{i} = {\bf v}_i] \Big );
\label{eq:rewrite}
\end{equation}
for example, suppose $Y \in \{0,1\}$, $E_{j}$ consists of 100 cases of type `A,' which $j$ treats with probability .75, for which $Y_{i}(j) - Y_{i}(j^{a}) = 1$, and 100 cases of type `B,' which $j$ treats with probability .01, for which $Y_{i}(j) - Y_{i}(j^{a}) = -1$. In line with the assessment practices typically employed by institutions, which evaluate agents only on the cases they were assigned, we restrict attention to measures defined over the distribution of case covariates in ${\cal I}_j$.

The disparities above measure whether agent $j$ behaves similarly or not to her peers on average, but do not measure the magnitude of the differences between $j$ and her peers, which may be quite large even if (\ref{eq:D1}) and (\ref{eq:D2}) are small.  Thus, we also consider a measure of the average size of the disparities, 
\begin{equation}
D_{3}(j) = N_{j}^{-1}\sum_{i \in {\cal I}_{j}} \big | \mathbb{E}[Y_{i}(j) - Y_{i}(j^{a}) \mid {\bf V}_{i} = {\bf v}_i] \big |.
\label{eq:D3}
\end{equation}
It is important to note that (\ref{eq:D3}) is an average of the absolute value of the expected disparities, not the average of the expected absolute disparities, $|Y_{i}(j) - Y_{i}(j^{a})|$. The two are related by the inequality $D_{3}(j) \leq N_{j}^{-1}\sum_{i \in {\cal I}_{j}}\mathbb{E}[|Y_{i}(j) - Y_{i}(j^{a})|\mid {\bf V}_{i} = {\bf v}_i].$ We revisit this point in the subsequent section on agreement.

The local disparities above compare the performance of an agent with her peers, but do not measure institutional performance. To do so, we construct the global measure
\begin{equation}
D_{3}(\cdot) = \sum_{j \in {\cal J}}\frac{N_{j}}{N} D_{3}(j), 
\label{eq:D1s}
\end{equation}
the average of the disparities $D_{3}(j)$, weighted by caseload.

\subsubsection*{Identification}
Without additional assumptions, the disparities above are not identified 
\citep[page 62]{van2000asymptotic}. We consider identifying assumptions here, discussing additional assumptions for estimation subsequently.

Let $Z_{i}^{a}$, with values $j^{a} \neq j$, denote an alternate agent selected for comparison with $j$.  The prima facie disparity between agent $j$ and her peers is then defined as
\begin{equation}
D_{1P}(j)= N_{j}^{-1}\sum_{i \in {\cal I}_{j}} \big (\mathbb{E}[Y_{i} \mid Z_{i}= j, \, {\bf V}_{i} = {\bf v}_i] - 
\mathbb{E}[Y_{i} \mid Z_{i}^{a} = j^{a}, \, {\bf V}_{i} = {\bf v}_i] \big ).  
\label{eq:disparity3}
\end{equation}
We state assumptions under which $D_{1P}(j) = D_{1}(j)$: 

\begin{itemize}
\item[{\bf A3}.] Positivity (Eligibility): For all $i \in {\cal I}$ and $j \in {\cal J}_{i}$, $0 < \Pr(Z_{i} = j \mid {\bf V}_{i} = {\bf v}_i) < 1$.
\item[{\bf A4}.] Unconfoundedness: For all $i \in {\cal I}$,
\begin{equation}
\{Y_{i}(j), Y_{i}(j^{a}) \}_{j, \, j^{a}\in {\cal J}_{i}, \ j \neq j^{a}} \perp\!\!\!\perp (Z_{i},Z_{i}^{a}) \mid {\bf V}_{i}, 
\label{eq:A4}
\end{equation}
where the symbol $\perp\!\!\!\perp$ denotes statistical independence.
\end{itemize}

Under {\bf A1}-{\bf A3}, $\mathbb{E}[Y_{i} \mid Z_{i} = j, \, {\bf V}_{i} = {\bf v}_{i}]$ is nonparametrically identified for each 
case $i \in {\cal P}$ and every agent $j \in {\cal J}_{i}$. Under the additional assumption {\bf A4}, $\mathbb{E}[Y_{i} \mid Z_{i}= j, \, {\bf V}_{i} = {\bf v}_{i}] = 
\mathbb{E}[Y_{i}(j) \mid {\bf V}_{i} = {\bf v}_{i}]$, $j \in {\cal J}_{i}$.  Thus, under {\bf A1}-{\bf A4}, $D_{1}(j)$, $D_{3}(j)$ and $D_{3}(\cdot)$  are non-parametrically identified.

\subsection{Agreement}
When a community standard is used to monitor the performance of agents, the agreement between agents' outcomes is of interest. If agent $j$ tends to disagree with her peers, further scrutiny of her behavior may be warranted. Widespread disagreement among agents may suggest institutional failure. Thus we construct local and global measures of agreement analogous to the disparities in Section 2.1. 

Disparity measures are often used inappropriately to infer the agreement among agents. For example, \citet{ramji2007} (hereafter RN), and \citet{GAO2016} use the disparities $D(j,j^{\prime})$ to infer widespread disagreement among federal immigration judges. Indeed, RN title their paper ``Refugee Roulette: Disparities in Asylum Adjudication and Proposals for Reform.'' Karen Musalo, another leading scholar, states ``whether you win or lose should not depend on the roll of the dice of which judge gets your case'' \citep{reuters2017}. But such inferences are unwarranted, even in a completely randomized experiment. Disparity measures depend on the marginal distributions of agents' potential outcomes, but agreement depends on the joint distribution. The marginal distributions may provide little information about the joint: for example, on a binary scale, two agents, each of whom grants asylum 50\% of the time, 
will agree  50\% of the time under independence, but may never agree or always agree if their classifications are dependent.

\subsubsection*{Measures for Categorical Outcomes}

For data measured on a nominal scale with $C$ categories, an obvious measure of agreement between two raters is $p^{o}$, the percentage of cases on the diagonal of the $C \times C$ cross-classification table. In a landmark paper that shaped the subsequent literature, Cohen (1960) rejected the use of $p^{o}$ and proposed the measure $\kappa = (p^{o} - p^{e})/(1 - p^{e}) \in [-1,1],$ in which $p^{e}$ is the ``chance agreement'' estimated under the assumption that the two sets of ratings are independent. Many generalizations of $\kappa$, to three or more raters, also to ordinal data, have since been proposed \citep{gwet2014}.

Despite its widespread use with categorical data, we do not find $\kappa$ or variants that adjust for chance agreement useful for measuring the performance of public institutions and their agents. If two agents independently classified each of $K$ cases, the joint distribution of outcomes would only be independent if the cases did not have common features that one or both agents used in their classifications. This is unlikely in real applications, and thus it is not clear why 
chance agreement should be the yardstick against which to measure actual agreement. Moreover, performance rankings based on $\kappa$ do not necessarily reflect rankings based on $p^0$, as $\kappa_{1} < \kappa_{2}$ does not imply $p_{1}^{o} < p_{2}^{o}$.  

Similarly, while models for the association between agents may be useful for estimating agreement \citep{becker1989using}, 
measures of association are not indicative of agreement per se \citep{bishop1975discrete}. Agreement depends on the marginal distributions of outcomes of different agents, whereas association does not, when measured by the odds ratio or functions of the odds ratio.  Consider the two by two table with row margins .9 and .1 and column margins .1 and .9., where the agents agree 20\% of the time when the odds ratio is infinite, whereas in the two by two table with homogeneous margins .9 and .1, 82\% of the cases lie on the diagonal under independence.

Let $\Pr(Y_{i}(j) = y, \, Y_{i}(j^{a}) = y^{\prime} \mid {\bf V}_{i} = {\bf v}_i) \equiv
\pi_{i}^{y,y^{\prime}}(j,j^{a})$
denote the probability agent $j \in {\cal J}_{i}$ responds
$y$ to case $i$ and agent $j^{a}$ responds $y^{\prime}$, and 
let 
$\sum_{y}\pi_{i}^{y,y}(j,j^{a}) \equiv  A_{i}^{=}(j,j^{a})$ 
denote the probability $j$ and $j^{a}$ agree, where $j^{a}$ is randomly selected from ${\cal J}_{i} \setminus j$ for each case $i$ as in Section 2.1. Let 
$A_{i}^{y \mid y}(j^{a} \mid j) \equiv$
$\Pr(Y_{i}(j^{a}) = y \mid Y_{i}(j) = y, {\bf V}_{i} = {\bf v}_i)$. 

To compare agent $j$ with her peers on the cases she was assigned, two measures are of primary interest, the probability $j$ agrees with peers,
\begin{equation}
A_{1}^{=}(j) = N_{j}^{-1}\sum_{i \in {\cal I}_{j}}A_{i}^{=}(j,j^{a}),
\label{eq:A1L}
\end{equation}
and the conditional probability $j^{a}$ agrees with $j$, 
\begin{equation}
A_{1}^{y \mid y}(j) = N_{j}^{-1}\sum_{i \in {\cal I}_{j}}A_{i}^{y\mid y}(j^{a} \mid j).
\label{eq:CA1L}
\end{equation}
Global analogues of the local measures are also readily constructed,
\begin{equation}
A_{1}^{=}(\cdot) = \sum_{j \in {\cal J}}\frac{N_{j}}{N}A_{1}(j),
\label{eq:A1G}
\end{equation}
\begin{equation}
A_{1}^{y \mid y}(\cdot) = \sum_{j \in {\cal J}}\frac{N_{j}}{N}A_{1}^{y\mid y}(j) \\[1em]
\label{eq:CA1G}
\end{equation}

While our measures of disparity and agreement do not depend on the fact that $Y_{i}(j)$ and $Y_{i}(j^{a})$ are binary, in such cases, agreement and 
disparity are related by the identities \begin{align}
\label{eq:agree_disp_identity}
A_{i}^{=}(j,j^{a}) &= 1 - {\mathbb E}[|Y_{i}(j) - Y_{i}(j^{a})|\mid {\bf V}_{i} = {\bf v}_i] \notag \\
&= 1 - {\mathbb E}[(Y_{i}(j) - Y_{i}(j^{a}))^2 \mid {\bf V}_{i} = {\bf v}_i] \notag \\
&= 1 - D_{i1}(j,j^{a}) - 2 \pi_{i}^{0,1}(j,j^{a}),
\end{align}
where disparity $D_{i1}(j,j^{a}) \equiv {\mathbb E}[Y_{i}(j) - Y_{i}(j^{a}) \mid {\bf V}_{i} = {\bf v}_i]$
is identified from the marginal distributions 
$\pi_{i}^{y}(j) \equiv \Pr(Y_{i}(j) = y \mid {\bf V}_{i} = {\bf v}_i) = {\mathbb E}[Y_{i}(j) \mid {\bf V}_{i} = {\bf v}_i]$, and 
$\pi_{i}^{y}(j^{a}) \equiv \Pr(Y_{i}(j^{a}) = y\mid {\bf V}_{i} = {\bf v}_i) = {\mathbb E}[Y_{i}(j^{a})\mid {\bf V}_{i} = {\bf v}_i]$, but $\pi_{i}^{0,1}(j,j^{a})$ is not. Without further assumptions, the agreement is only ``partially identified'' by the FH bounds
\begin{align}
1 - D_{i1}(j,j^{a}) - 2 \min(\pi_{i}^{0}(j), \pi_{i}^{1}(j^{a})) &\leq A_{i}^{=}(j,j^{a}) \notag \\[1em] 
&\leq 1 - D_{i1}(j,j^{a}) - 2 \max(0, \pi_{i}^{0}(j) + \pi_{i}^{1}(j^{a}) - 1).
\label{eq:bound}
\end{align}
Bounds for measures (\ref{eq:A1L})-(\ref{eq:CA1G}) may be constructed from (\ref{eq:bound}), a point we revisit subsequently.

\subsubsection*{Measures for Continuous Outcomes}

Agreement between two raters $j$ and $j^{a}$ on a continuous scale $Y$ is commonly measured by the the intra-class correlation or concordance correlation coefficient,
\begin{equation}
\frac{2 \text{Cov}(Y(j),Y(j^{a}))}{\text{Var}(Y(j)) + \text{Var} (Y(j^{a})) + {\mathbb E}[Y(j) - Y(j^{a})]^{2}}
\label{eq:lin}
\end{equation}
(Lin 1989; Lin, Hedayat, Sinha and Yang 2002), motivating the local measure $L(j)$, 
\begin{equation}
N_{j}^{-1}\sum_{i \in {\cal I}_{j}}\dfrac{2 \text{Cov}(Y_{i}(j),Y_{i}(j^{a})\mid {\bf V}_{i} = \bf{v}_i)}{\text{Var} (Y_{i}(j) \mid {\bf V}_{i} = {\bf v}_i) + \text{Var} (Y_{i}(j^{a}) \mid {\bf V}_{i} = {\bf v}_i) + \mathbb{E}[Y_{i}(j) - Y_{i}(j^{a}) \mid {\bf V}_{i} = {\bf v}_i]^{2}}.
\label{eq:lin1}
\end{equation}
Global agreement can then be measured as the weighted average of the local measures, 
\begin{equation}
L(\cdot) =  \sum_{j \in \cal{J}}\frac{N_{j}}{N}L(j).
\label{eq:lin2}
\end{equation}

Equation (\ref{eq:lin}) is a natural extension of the second line of equation (\ref{eq:agree_disp_identity}). When Y is continuous and bounded, $m \leq Y \leq M$, we define 
\begin{equation}
A_{i}(j,j^{a}) \equiv 1 - \frac{{\mathbb E}[(Y_{i}(j) - Y_{i}(j^{a}))^2]}{(M-m)^2}, 
\label{eq:da}
\end{equation}
normalized here to take value 0 under maximal disagreement, and which reduces to the categorical case discussed in the previous section when the outcome is binary. When $Y$ is unbounded, observe that the numerator of (\ref{eq:da}) can be expanded as in $${\mathbb E}[(Y_1 - Y_2)^2] = \text{Var}(Y_{1}) + \text{Var}(Y_{2}) -2 \, \text{Cov}(Y_{1},Y_{2}) + ({\mathbb E}[Y_{1}] - {\mathbb E}[Y_{2}])^{2}.$$  
When the covariance is positive, $(M - m)^2$ can thus be replaced by the upper bound $c_2 = \text{Var}(Y_{1}) + \text{Var}(Y_{2}) + ({\mathbb E}[Y_{1}] - {\mathbb E}[Y_{2}])^{2}$, in which case (\ref{eq:da}) reduces to (\ref{eq:lin}).

Other extensions are possible. Using the first line of equation (\ref{eq:agree_disp_identity}) would produce the numerator ${\mathbb E}[|Y_{i}(j) - Y_{i}(j^{a}))|]$. The denominator would be $M-m$ in the bounded case, $c_1 = {\mathbb E}[|Y_{1}| + |Y_{2}|]$ in the unbounded case, or if second moments exist, $\sqrt{c_{2}}$ or $\min(c_{1}, \sqrt{c_{2}})$. Another approach is to replace the denominator with the FH bounds using the distribution functions $F_{Y_{1}}$ and $F_{Y_{2}}$. Measures of conditional agreement can be constructed analogously. Further consideration of the continuous case, including the partial identification of agreement measures when one agent is observed per case, will be taken up in future work. 

\subsubsection*{Identification}
When each case is handled by one agent, 
measures $A_1^=(j)$, $A_1^=(\cdot)$,  $L(j)$, and $L(\cdot)$ are identified up to the FH bounds. Point identification requires additional assumptions.

The simplest assumption is that outcomes are independent given observed covariates, $Y_{i}(j) \perp\!\!\!\perp Y_{i}(j^{a}) \mid {\bf V}_{i}$. Agreement is then identified from the marginal distributions.  But this assumption will be violated in the common scenario in which ${\bf V}_{i}$ fails to include relevant but unrecorded information that both $j$ and $j^{a}$ would see in handling case $i$, where ${\bf U}_{i}(j) = {\bf U}_{i}(j^{a}) \equiv {\bf U}_{i}$. 
For example, observed covariates rarely capture the weight of the evidence an asylum seeker might present at trial to support a fear of persecution. But even though ${\bf U}_{i}$ is unobserved, consideration of the way $({\bf V}_{i},{\bf U}_{i})$ is likely to impact the joint distribution of outcomes is useful for tightening the FH bounds. 

We begin by strengthening the unconfoundedness assumption ${\bf A4}$:
\begin{itemize}
\item[{\bf A4*}.] Unconfoundedness: For all $i \in {\cal I}$,
\begin{equation}
\{Y_{i}(j), Y_{i}(j^{a})\}_{j,\, j^{a} \in {\cal J}_{i}, \ j \neq j^{a}}, {\bf U}_{i} \perp\!\!\!\perp (Z_{i},Z_{i}^{a}) \mid {\bf V}_{i}. 
\label{eq:A3*}
\end{equation}
\end{itemize}

Assumption  {\bf A4*}  implies unconfoundedness given $({\bf V}_{i},\bf{U}_{i})$ and also unconfoundedness given ${\bf V}_{i}$ (Assumption {\bf A4}). In other words, for each case $i$, ${\bf U}_{i}$ is a vector of ``unconfounded unmeasured predictors'', not an unobserved confounder. 

We also assume that if agents $j$ and $j^{a}$ observed case $i$, with information $({\bf V}_{i}, {\bf U}_{i})$, each agent would evaluate the case independently of the other agent, and the two agents would process the additional information in ${\bf U}_{i}$ in a similar manner:

\begin{itemize}
\item[\label{independence}{\bf A5}.]   Independence of Agents Outcomes with Respect to Recorded and Unmeasured Information: For all $i \in {\cal I}$,  
\begin{equation}
Y_{i}(j) \perp\!\!\!\perp Y_{i}(j^{a}) \mid {\bf V}_{i}, \, {\bf U}_{i},
\label{agent}
\end{equation} 
\end{itemize}

\begin{itemize}
\item[{\bf A6}.] Common Interpretation of Unmeasured Information: 
If 
\begin{align*}
\mathbb{E}[Y_{i}(j) \mid {\bf V}_{i} &= {\bf v}, {\bf U}_{i} = {\bf u}] \geq \mathbb{E}[Y_{i}(j) \mid {\bf V}_{i} = {\bf v}, {\bf U}_{i} = \bf{u}^{*}], \quad \text{then} \\
\mathbb{E}[Y_{i}(j^{a}) \mid {\bf V}_{i} &= {\bf v}, {\bf U}_{i} = {\bf u}] 
\geq \mathbb{E}[Y_{i}(j^{a})\mid {\bf V}_{i} = {\bf v}, {\bf U}_{i} = {\bf u}^{*}].
\end{align*}
\end{itemize}

\subsubsection*{A Covariance Bound} 
Many measures, including $A_1^=(j)$, $A_1^=(\cdot)$,  $L(j)$, $L(\cdot)$, depend on the covariances $\text{Cov}(Y_{i}(j),Y_{i}(j^{a}) \mid {\bf V}_{i} = {\bf v}_i)$. For example, in the binary case, (\ref{eq:agree_disp_identity}) implies
\begin{align}
A_{i}^{=}(j) &= N_{j}^{-1}\sum_{i \in {\cal P}_{j}}\pi_{i}^{=}(j,j^{a}) \nonumber \\
&= N_{j}^{-1} \sum_{i \in {\cal P}_{j}} \Big [1 - \pi_{i}^{y}(j) -  \pi_{i}^{y}(j^{a}) + 2 \, \pi_{i}^{y}(j) \, \pi_{i}^{y}(j^{a}) + 2 \, \text{Cov}(Y_{i}(j),Y_{i}(j^{a})\mid {\bf V}_{i} = {\bf v}_i ) \Big ]
\nonumber \\
&= N_{j}^{-1} \sum_{i \in {\cal P}_{j}}\Big [1 - \pi_{i}^{y}(j) -  \pi_{i}^{y}(j^{a}) + 2 \, \pi_{i}^{y}(j) \, \pi_{i}^{y}(j^{a}) \Big ] + 2 \, \mathbb{E}\left [\text{Cov}(Y_{i}(j),Y_{i}(j^{a})\mid {\bf V}_{i}) \right ].
\label{eq:cov}
\end{align}
The expectation is taken over the empirical distribution of cases assigned to agent $j$. 

We now use assumptions
{\bf A5} and {\bf A6} to bound the unidentified covariances, and thus the agreement measures. 
We begin with the covariance between agent outcomes, conditioned on the observed information, which can be written
 \begin{align}
\label{eq:cov-UV-judges}
\text{Cov}\bigl(Y_i(j),\,Y_i(j^a) &\mid {\bf V}_{i}\bigr)
=
\mathbb{E}\Big[
\text{Cov}\bigl(Y_i(j),\,Y_i(j^a) \mid {\bf U}_{i}, {\bf V}_{i}\bigr) \,\Big|\, {\bf V}_{i}
\Big] \notag \\
&+ 
\text{Cov}\Big(
\mathbb{E}[Y_i(j) \mid {\bf U}_{i}, {\bf V}_{i}], \,
\mathbb{E}[Y_i(j^a) \mid {\bf U}_{i}, {\bf V}_{i}]
\;\Big|\;
{\bf V}_{i}
\Big).
\end{align}
By {\bf A5}, the first term on the right hand side of (\ref{eq:cov-UV-judges}) is 0.  
Furthermore, {\bf A6} implies the second term is positive, $\text{Cov}\Big(
\mathbb{E}[Y_i(j) \mid {\bf U}_{i}, {\bf V}_{i}], \,
\mathbb{E}[Y_i(j^a) \mid {\bf U}_{i}, {\bf V}_{i}]
\;\Big|\;
{\bf V}_{i}
\Big) \, \geq \, 0.$
It follows that $\text{Cov}(Y_{i}(j),Y_{i}(j^{a})\mid {\bf V}_{i} = {\bf v}) \geq 0$. Thus, in the categorical case, $A_{i}^{=}(j,j^{a}) \geq \sum_{y} \pi_{i}^{y}(j) \, \pi_{i}^{y}(j^{a})$.

To obtain upper bounds that are tighter than the FH bounds, we again use the law of total covariance, which, under {\bf A5}, yields
\begin{align}
\label{eq:cov2}
\text{Cov}(Y_{i}(j),Y_{i}(j^{a})) = 
{\mathbb E} \Big [ \text{Cov} \big (
\mathbb{E}[Y_i(j) \mid {\bf U}_{i}, {\bf V}_{i}], \,
\mathbb{E}[Y_i(j^a) \mid {\bf U}_{i}, {\bf V}_{i}]
\mid
{\bf V}_{i}
\big ) \Big ] \nonumber \\ 
+ \, \text{Cov}\Big(
{\mathbb E}[Y_i(j) \mid {\bf V}_{i}], \,
{\mathbb E}[Y_i(j^a) \mid {\bf V}_{i}] \Big),
\end{align}
a decomposition of the covariance as the sum of a ``within'' ${\bf V}_{i}$ component measuring the added covariation attributable to the unrecorded information ${\bf U}_{i}$ and a ``between'' component measuring the covariation between agents expectations, given the observed information ${\bf V}_{i}$.

Of interest here is the unidentified ``within component''.  However, the ``between'' covariance is identified, and if positive, as when agents $j$ and $j^{a}$ react similarly to observed conditions, it can serve as a yardstick for the unidentified within covariance, which can be writen as 
a multiple $\alpha \geq 0$ of the ``between'' component. We can then replace the within component by this multiple of the between component under various assumptions about the value of $\alpha$. For example, if $\alpha = 0$, the information in ${\bf U}_{i}$ adds no predictive value net of ${\bf V}_{i}$; if $\alpha = 1$, the predictive value of the unrecorded information is of equal magnitude to the predictive value of the recorded information; and so on. As the net contribution of the information in ${\bf U}_{i}$ increases, $\alpha$ increases, pushing agreement toward the upper FH bound.

\subsection{Estimation}

\subsubsection*{Disparity} 
Our goal is to estimate the case-level expected outcomes ${\mathbb E}[Y_{i}(j) \mid {\bf V}_{i}]$ and ${\mathbb E}[Y_{i}(j^{a}) \mid {\bf V}_{i}]$ for the set $i \in {\cal I}_j$, from which the estimation of the disparity measures follow.  
While the estimation of ${\mathbb E}[Y_{i}(j) \mid {\bf V}_{i}]$ is straightforward for $i \in {\cal I}_j$ under assumptions previously made, estimation of ${\mathbb E}[Y_{i}(j^{a}) \mid {\bf V}_{i} = {\bf v}_i]$ for $i \in {\cal I}_j$ must be constructed from estimates of ${\mathbb E}[Y_{k}(j^{a}) \mid {\bf V}_{k} = {\bf v}_i]$ for $k \in {\cal I}_{j^{a}}$, necessitating additional assumptions:

\begin{itemize}
\item[{\bf A7}.] Independent Sampling: The random data vectors ($Q_{1},...,Q_{N})$ are statistically independent, where $Q_{i} = (\{Y_{i}(j)\}_{j \in {\cal J}_{i}}, Z_{i}, Z_{i}^{a}, {\bf V}_{i})$,

\item[{\bf A8}.] Intra-Agent Consistency: 
For every $i$, $j^{a} \in {\cal J}_{i} \setminus j$, and $k \in {\cal I}_{j^{a}}$, $$\mathbb{E}\left[Y_{i}(j^{a}) \mid {\bf V}_{i} = {\bf v}, \, {\bf U}_{i} = {\bf u}\right] = \mathbb{E}\left[Y_{k}(j^{a}) \mid {\bf V}_{k} = {\bf v}, \, {\bf U}_{k} = {\bf u}\right].$$
\end{itemize}

Assumption \textbf{A8} justifies using
the outcomes of cases $j^{a}$ was assigned to estimate the outcomes of cases $j^a$ was eligible for but not in fact assigned. We assume agent $j^{a}$ would treat both sets of cases the same, provided the same information {\bf u} and {\bf v} were observed.

A final assumption is necessary because ${\bf U}$ is not observed, and we can only estimate $\mathbb{E}\left[Y_{i}(j) \mid {\bf V}_{i} = {\bf v}_i\right]$ and $\mathbb{E}\left[Y_{k}(j^{a}) \mid {\bf V}_{k} = {\bf v}_i\right]$. 

\begin{itemize}
\item[{\bf A9}.] Exchangeability of Unmeasured Characteristics: For all $i \in \cal{I}$, ${\bf U}_{i} \mid {\bf V}_{i} = {\bf v}$ are independently and identically distributed. 
\end{itemize}
Assumptions \textbf{A8} and \textbf{A9} jointly imply $\mathbb{E}\left [Y_{i}(j^{a}) \mid {\bf V}_{i} = {\bf v} \right ] = \mathbb{E} \left [Y_{k}(j^{a}) \mid {\bf V}_{k} = {\bf v}\right ]$.   

\subsubsection*{Agreement} 
In Section 2.2, we proposed a covariance bound for agreement measures such as $A_{i}^{=}(j)$ and $A_{i}^{=}(\cdot)$, which are partially identified from the marginal expectations ${\mathbb E}[Y_{i}(j) \mid {\bf V}_{i}]$, ${\mathbb E}[Y_{i}(j^{a}) \mid {\bf V}_{i}]$. 
Having estimated the marginal expectations under assumptions such as {\bf A1}-{\bf A9}, one can estimate  agreement using various $\alpha$ values, replacing $\text{Cov}(Y_{i}(j), \,  Y_{i}(j^{a}) \mid {\bf V}_{i})$ with $\alpha \ \text{Cov}(\mathbb{E} \left [ Y_{i}(j) \mid {\bf V}_{i} \right ], \, \mathbb{E} \left [Y_{i}(j^{a}) \mid {\bf V}_{i} \right ] )$ in equation (\ref{eq:cov}). If the information in ${\bf U}_{i}$ is irrelevant, given ${\bf V}_{i}$, $\alpha = 0$ and the agents' classifications are independent. As $\alpha$ increases, the value of the information in ${\bf U}_{i}$ increases, pushing agreement toward the upper FH bound. It is important to note that when we estimate agreement between judges using different values of $\alpha$, we always threshold the agreement at the upper FH bound.

A related approach is to pretend 
${\bf V}_{i} = ({\bf V}_{i1},{\bf V}_{i2})$ contains all the relevant information and to compute the relative contribution $\alpha ({\bf V}_{i2};{\bf V}_{i1})$
when ${\bf V}_{i2}$ is omitted. Both approaches are similar in spirit to sensitivity analyses in causal inference that ask how strong the association between confounders and outcomes and confounders and treatment must be to drive a causal effect to 0 (e.g.,
\cite{cinelli2020}). But here we are not attempting to minimize or maximize agreement, and as
${\bf U}_{i}$ is not a confounder for case $i$ only the additional association induced by ${\bf U}_i$ net of ${\bf V}_i$ is of interest.

\section{Application: Disparity and Disagreement in the U.S. Asylum Courts}
Under the Geneva Convention of July, 28 1951 and its 1967 protocol, the United States, a signatory, is required to grant asylum to an applicant who arrives at a port of entry or is inside the USA, and qualifies as defined in the Refugee Act of 1980 as a person ``unwilling or unable to return to'' their home country due to a ``well founded fear of persecution based on race, religion, nationality, membership in a particular social group, or political opinion.''

Two types of asylum cases end up in the Executive Office for Immigration Review (EOIR) courts.  Affirmative applicants first petition the United States Citizenship and Immigration Service (USCIS), where an asylum officer conducts a ``merits'' review and decides whether to grant asylum to the applicant. Removal proceedings for unsuccessful applicants without lawful status are then initiated with a notice to appear (NTA) in court before an EIOR judge for a ``master calendar'' hearing.   Persons who did not apply affirmatively and who have been apprehended by U.S. authorities also receive an NTA and may apply ``defensively,'' as may persons in expedited removal proceedings who demonstrate to the satisfaction of the asylum officer a credible fear of torture or persecution if returned to their home country.  

Recipients of the NTA who fail to attend the master calendar hearing are typically issued a removal order ``in absentia.'' Those who attend are asked to admit to or deny the allegations in the NTA and to state what type of relief is sought, if any.  For persons who request relief (e.g., asylum), the judge then schedules a merits hearing over which he/she normally presides.   Some applicants withdraw their request prior to the hearing. The remaining applicants present evidence to support their request at the hearing, after which the judge rules on their application.  Of the 56,898 affirmative applications filed in 2014, 14,758 applicants were granted asylum, and of the more than 30,000 defensive applicants, 8,775 were granted asylum \citep{mossaad2016}.

\subsection{A Review of Two Influential Studies}
`Large' disparities in asylum grant rates within and across courts have been reported. \citep{GAO2016, miller2014, ramji2007, TRAC183, TRAC490, TRAC590, TRAC670}.
We examine two prominent studies to illustrate the evidence
immigration researchers and policy advocates have used to conclude the EOIR judiciary lacks impartiality.  

\subsubsection*{Example 1: Refugee Roulette}
The dataset analyzed by RN contain 269,756 EOIR cases.  After excluding ``cases that were abandoned, withdrawn or disposed of in some other way'' (p. 396), 140,428 EIOR ``merits'' cases remained. Subsequent exclusions include affirmative applications filed by Mexican nationals, who the authors claim were primarily seeking other forms of relief, and most applicants held in detention.  
Cases with multiple family members were bundled together under a  ``lead'' case and treated unitarily, as asylum is typically granted (or not) to the lead and all secondary (rider) cases bundled with the lead.  

RN then analyze almost 80,000 cases that were decided between 2000 and 2004, focusing on seventeen ``high volume'' courts in which more than 1,500 applications per court were handled during the study period, and to applicants from fifteen nationalities where more than 500 cases per nationality were filed during 2004.  They view large cross-court differences, for example that between Orlando and Atlanta, with grant rates of 76\% and 7\%, respectively, for Chinese applicants, as evidence that judges do not behave impartially.

Substantial differences were also observed within courts, where judges are randomly assigned to cases.  In Los Angeles, among the 23 judges who heard 50 or more cases with Chinese applicants, grant rates ranged  from 9\% to 81\% for a disparity $\text{D}(j,j^{\prime}) = .72$ (.50 excluding the lowest and highest rates), and among the 23 judges in Miami who heard 50 or more cases from Columbian applicants, the grant rates ranged from 5\% to 88\% for a disparity $\text{D}(j,j^{\prime}) = .83$ (.71 excluding the lowest and highest rates). In Miami, Los Angeles and New York, approximately one third of the judges deviate from the court-wide average by more than 50\%. RN conclude (p. 296):``In many cases, the most important moment in an asylum case is the instant in which a clerk randomly assigns an application to a particular asylum officer or immigration judge''.  

RN also use logistic regression with a somewhat different sample of 66,443 affirmative cases to study the relationship of outcomes to judge and case characteristics. Applicants with legal representation, dependents, and who came from countries with a low national freedom index were more likely to obtain asylum. A judge's prior work experience and gender, with female judges more likely to grant asylum, were also associated with outcomes.  Given the large sample size, several other predictors were statistically significant, but with substantively insignificant coefficients. 

\subsubsection*{Example 2: GAO 2016}

The GAO analyze cases in 28 ``high volume'' courts that were decided between 1995 and 2014.  Like RN, applications from ``low volume'' countries, from judges who heard fewer than fifty affirmative cases and fifty defensive cases in the court they primarily served were excluded, and applications with multiple family members were bundled. But whereas RN excluded applications that were abandoned, withdrawn, or otherwise closed, the GAO treated these as denials. Another important difference is that GAO excluded cases from applicants in detention. Cases where the applicant or the government appealed the judge's decision were also excluded.

In an earlier study focusing on the period before May 2007, the GAO (2008) found substantial variation in grant rates among nationalities, courts, and judges. For the new period, May 2007 through fiscal year 2014, among affirmative applications, grant rates varied from  less than 5\% in Atlanta and Omaha to 66\% in New York, from 0 to 31.3\% among judges in Los Angeles, and from 22.9\% to 83.3\% among judges in New York.  Similar results, with somewhat lower grant rates, were reported for defensive applications. 

For each of four time periods and separately for defensive and affirmative applications, the GAO also estimated logistic-normal mixed effects models.  
Applicants with dependents and/or legal representation who applied affirmatively between May 2007 and the end of fiscal year 2014 were more likely to obtain asylum;
RN reported similar results.  Female judges and judges appointed by Bush senior were more likely to grant asylum than judges appointed by Bush junior or Obama, and years of service was negatively associated with success; 95\% confidence intervals for variances of random effects for judges, courts, and nationality and language of the asylum seeker were all bounded away from 0.  

The GAO also attempted to use these models to account for inconsistency between judges by constructing a 95\% prediction interval for a `representative applicant' who comes before two judges with the same characteristics, but different judge random effects.  For the 2007-2014 period, the upper and lower endpoints differ by 47 and 57 percentage points, respectively, for affirmative and defensive applicants. Were judge characteristics not held constant, presumably the difference would be even larger.

\subsection{Reanalysis of Disparity and Disagreement}

The validity of our analyses depends on satisfying the assumptions in Section 2. To do so, substantial departures from the analyses in RN and GAO are required.

Recipients of an NTA typically learn the identity of the judge assigned to their case prior to the master calendar hearing.  Affirmative applicants apply for relief prior to assignment. The remaining recipients apply defensively (if at all) after the judge is assigned and are thus of two types. One type would apply regardless of the judge assigned, while the other may choose to apply depending on the judge assigned to the case. These two types cannot be differentiated from the data and therefore, 
unlike RN and the GAO, we only consider affirmative applications. RN also excluded cases that were abandoned, withdrawn, or otherwise closed. But while the judge assignment could affect whether or not a defensive application is filed, affirmative applications are filed before a judge is assigned. 
Therefore, like GAO, we classify these outcome as ``not granted.''  

To justify interpreting the measures 
(\ref{eq:disparity1}) and (\ref{eq:disparity2}) 
as evidence of disparity and inconsistency, RN cite the random assignment, within courts, of judges to cases.
But 
this interpretation is unwarranted 
because some judges may be more likely 
to be assigned certain types of cases, e.g., cases with
minors.
In addition, judges do not necessarily serve for the entire study period, and the composition of cases may also vary over time. In fact,
two judges with non-overlapping service periods are not even eligible to hear any of the same cases.  
We avoid this problem by comparing the judge randomly assigned to case $i$ only to other judges who could have been assigned that case. Consistent with this, and unlike RN and GAO and others who interpret ``low'' grant rates in some courts, as compared to others, as evidence of judicial bias, we do not make cross-court comparisons, as such interpretations are also consistent with other explanations, e.g., differences in the composition of cases.

Further, RN and GAO fail to guard against post-treatment confounding.  Both analyze cases decided during their study periods.
These include difficult cases filed long before the study period began and cases filed toward the end of the period that were quickly resolved.  To avoid any potential problems due to non-ignorable censoring, we analyze cases filed (as versus decided) during our study period. They also associate outcomes to the judge who closed the case, who might be different than the judge initially assigned. This is problematic if reassignments are not randomized. But even if they were, it is not clear that an analysis using the closing judges would be preferred:
outcomes for cases $i$ reassigned to judge $j$ may not be indicative of those that would have occurred had $j$
been initially assigned because actions taken by the initially assigned judge may shape the course of future proceedings \citep{sobel2023association}.
They also note that applicants with legal representation were more likely to obtain asylum, and in their models of asylum outcomes, representation is included as a predictor. We do not include this predictor: in our data the decision to obtain representation or not was made for more than 10\% of the cases after the master calendar hearing, and may have depended on the judge assigned.  

Finally, even if all judges see similar cases on average, the largest within court disparities $D(j,j^{\prime})$, used by RN and GAO as evidence of judicial inconsistency, may not be indicative of more typical disparities or levels of agreement between judges.   
\subsubsection*{Data} Our study period, May 1, 2007 to September 30, 2014, spans the same interval as period 4 of the GAO study. While GAO studied cases decided during this period we study affirmative applications submitted during this period to avoid potential biases due to censoring.
Like RN and GAO, groups of individuals that applied together are treated as a single unit. During this period, 122,235 applications were filed. We consider the 
subset of 120,483 applications where the petitioner had not previously applied for asylum. See Section A of the Supplement for additional information on the construction of the data set used.

In 2014, there were 59 immigration courts. During our study period, more than half the cases were first heard in the top three courts (New York City (34.5\% of cases), West Los Angeles (20.45\%),  
San Fransisco (5.3\%), and 90\% of cases were heard at the top 16 locations: as we model outcomes court by court, attention is limited to these locations. 

Each case $i$ is assigned to a judge. 
In 28.7\% of cases, different judges presided over the first and last proceeding. We cannot account for the re-assignment mechanism. As in `intent to treat' analyses, we associate the outcome of case $i$ to the judge first randomly assigned to the case. We must also select a judge $j^{a}$ who could have been assigned case $i$ but was not. Judges are typically assigned to cases after the NTA, which specifies the date of the master calendar hearing, has been issued: we assume $j^{a}$ is chosen from the judges scheduled to be in court on that date. 
An alternate judge was available for 92,218 (of 102,388 cases in the top 16 courts); cases without alternates were omitted from analyses. 
For additional information on the 16 courts and the judges in these courts see Section B of the Supplement.

We use Bayesian models to compare assigned judges with their alternates. We do not attempt to estimate models for individual judges. Thus, in addition to case covariates ${\bf V}_{i}$, to account for heterogeneity between judges, we include in our models covariates ${\bf W}_{j}$ that have been shown to be predictive of outcomes in previous work. 
Like GAO, we include cases that were granted (29.6\%), denied (10.3\%), abandoned (6.4\%), withdrawn (17.5\%) or other (36.2\%). Unlike GAO, we do not exclude cases where the judge's decision was appealed: the outcome $Y_{i}=1$ if case $i$ resulted in a grant of asylum, 0 otherwise. 

The case covariates ${\bf V}_{i}$ are: 
1) RIDERS, 1 if there are riders attached to the case, 0 otherwise; 
2) N\_RIDERS, the number of riders;
3) ENGLISH, 1 if applicant speaks English, 0 otherwise, 4) SPANISH, 1 if applicant speaks Spanish, 0 otherwise, 5) CHINA, 1 if applicant is Chinese, 0 otherwise,  5) LANGUAGE, primary language spoken by the applicant, 6) NATIONALITY, the applicant's nationality;
7) APP\_YEAR, the year the asylum application was filed;
8) ON\_TIME, a categorical variable coded 0 if the applicant's arrival time is unknown, 1 if the arrival time is known and application was made within one year of arrival in the US, 2 otherwise;
9) NTA\_YEAR, the year the notice to appear was sent;
10) HEARING\_DATE, the date (year-month-date) the hearing was held.

The judge covariates ${\bf W}_{j}$ are: 1) SEX, 1 if male, 0 otherwise;
2) WORK1, 1 for previous work experience with Immigration and Naturalization Services or Immigration and Customs Enforcement, 0 otherwise;  3) WORK2, 1 for previous work experience with the Department of Justice, 0 otherwise, 4) MILITARY, 1 for prior military service, 0 otherwise;
5) YEAR\_BA, year the judge received the bachelor's degree;
6) YEAR\_JD, year the judge received the law degree;
7) YEAR\_AP, year the judge was appointed.

\subsubsection*{Model and Estimation} 

For each court, we model the marginal probability that case $i$ would receive asylum under judge $j$,
\[
  \Pr\{Y_i(j)=1\mid \mathbf V_i, \mathbf W_j\}
  =
  F\{G(\mathbf V_i, \mathbf W_j, \, j)\},
\]
where 
$G$ is an unknown prediction function and $F$ is a link function such as the probit, the standard normal cumulative distribution function.

We estimate the marginal probabilities using a variant of Bayesian Additive Regression Trees \citep{chipman2010bart}, hereafter BART, in which the prediction function is represented as a sum of $m$ regression trees: 
\[G(\mathbf V_i, \mathbf W_j, \, j) = \sum_{t=1}^{m} g(\mathbf V_i, \mathbf W_j, \, j; \ T_t,M_t),
\] where $T_t$ is a binary regression tree that partitions the covariate space and $M_t$ is the set of terminal-node parameters associated with that tree. We anticipate the grant probability varies smoothly over time and use soft trees as described in \citet{linero2018bayesian}, setting $m=200$ trees.

Regularization is achieved by specifying a prior distribution on $(T_t, \, M_t)$, as described in \citet{linero2022softbart} (see also \citet{chipman1998bayesian}), which limits the influence of any individual tree. Specifically, we shrink the node-parameters to $0$ on the covariate scale, which shrinks the overall predictions to $0.5$ on the probability scale, which errs on the side of uncertainty over whether judges agree or disagree.

We note that nonparametric approaches to causal inference are increasingly used because they can flexibly capture nonlinearities and interactions without requiring these features to be specified in advance \citep{hill2011bayesian}. Thus, we estimate and report the results from a fully nonparametric probit BART model with the \texttt{R} package \texttt{SoftBart}
using covariates ${\bf V}_{i}$ and ${\bf W}_{j}$, and judge indicators $j$ as predictors. The \texttt{SoftBart} implementation mitigates some of the biases that can result in large samples \citep{li2023bayesian, antonelli23} and posterior consistency is achieved under fairly general conditions \citep{linero2018bayesian}. Nevertheless, we also compare our results with estimates obtained from several other Bayesian models and find similar results.
The first is a parametric logistic regression estimated with \texttt{brms}. It includes ${\bf V}_{i}$, ${\bf W}_{j}$, 
and random intercepts for nationality, language, and judge. The second is a semiparametric \texttt{stan4bart} model. In this specification, BART components are used for the  covariates ${\bf V}_{i}$ and ${\bf W}_{j}$, while judge-specific heterogeneity is represented by a judge random effect. We also compare these results with another fully nonparametric implementation of BART using the \texttt{R} package \texttt{BART} \citep{sparapani2021nonparametric}.

For each model, we draw 7,000 samples from the posterior distribution of the estimands using Markov chain Monte Carlo. We discard the first 2,000 warm-up iterations, retaining 5,000 draws for inference. The models produce similar posterior means, while the fully nonparametric model yields wider posterior distributions: we report the results from this model.
See Section C of the Supplement for additional model and estimation details, and Section D for presentation of the results using the other models.

\subsubsection*{Estimation of Disparities and Agreement}
For each of the 5,000 draws above, we estimate the marginal grant probability for the assigned judge and an alternate, chosen with equal probability from the set of eligible alternates, defined to be the set of judges that held a master calendar hearing on the same day as case $i$. We then plug these into the formulae from Section 2, producing estimates of disparity and agreement: posterior summaries and uncertainty intervals are calculated directly from these. Note that an alternate judge is chosen once for each case used in all posterior samples.

As discussed in Section 2, agreement, equation (\ref{eq:A1G}), is only partially identified, 
known only up to the covariance ``within'' ${\bf V}_{i}$. We replace this term with the ``between'' covariance 
times a sensitivity parameter $\alpha$ as described in Section 2.2. If a particular value of $\alpha$ under consideration produces a larger value of agreement for a case than that implied by the upper FH bound, we replace this larger value with the upper FH bound.

\subsubsection*{Results}
\begin{figure}
    \centering
    \caption{Judge disparities $D_1(j)$ and $D_{3}(j)$ in New York City (NYC, top) and West Los Angeles (WLA, bottom).}
    \label{fig:disparity}

    \vspace{1em}
        
    \begin{subfigure}{\textwidth}
        \centering
        \caption{NYC}
        \label{fig:disparity_top}
        \vspace{1em}
        \includegraphics[width=.9\linewidth,
                         trim= 0 1cm 0 .76cm, 
                         clip
                         ]{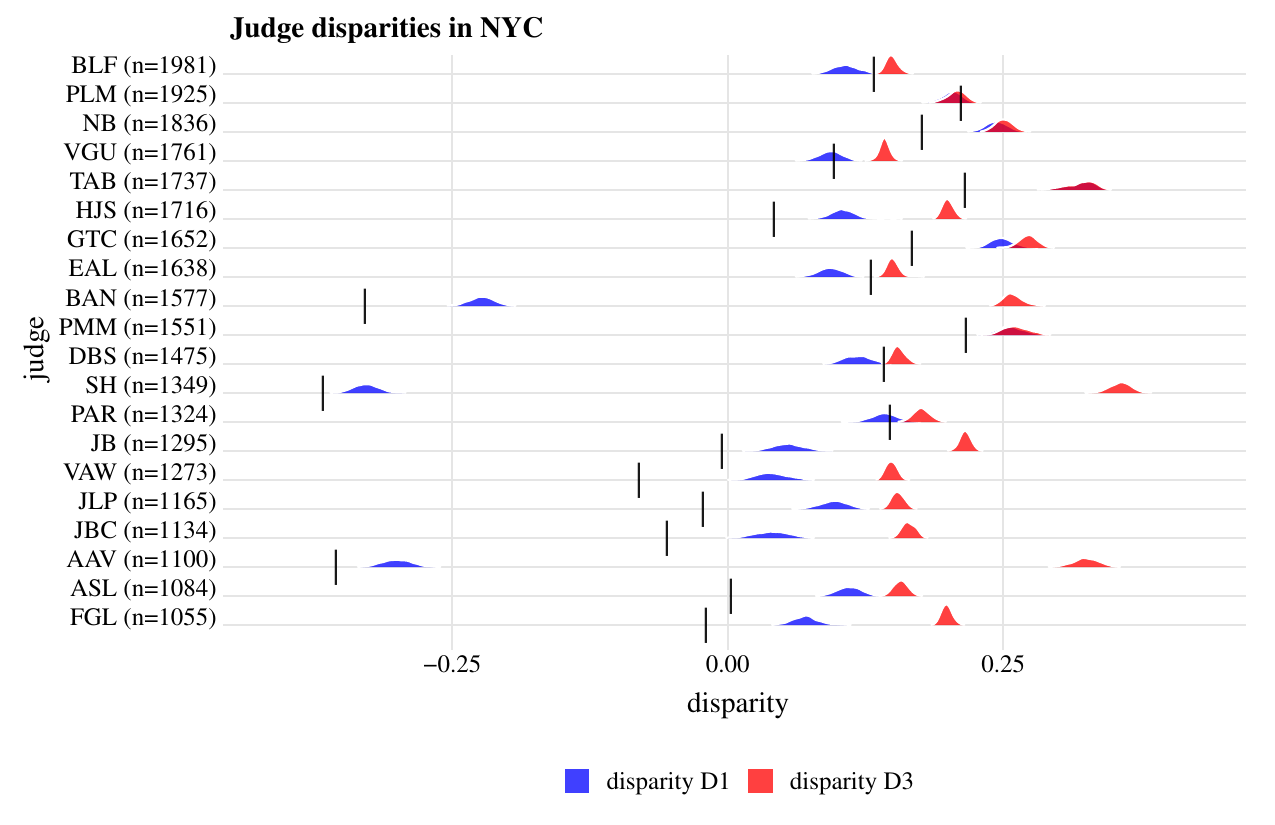}

    \end{subfigure}

    \begin{subfigure}{\textwidth}
        \centering
        \caption{WLA}
        \label{fig:disparity_bottom}
        \vspace{1em}
        \includegraphics[width=.9\linewidth,
                         trim= 0 0 0 .76cm, 
                         clip
                         ]{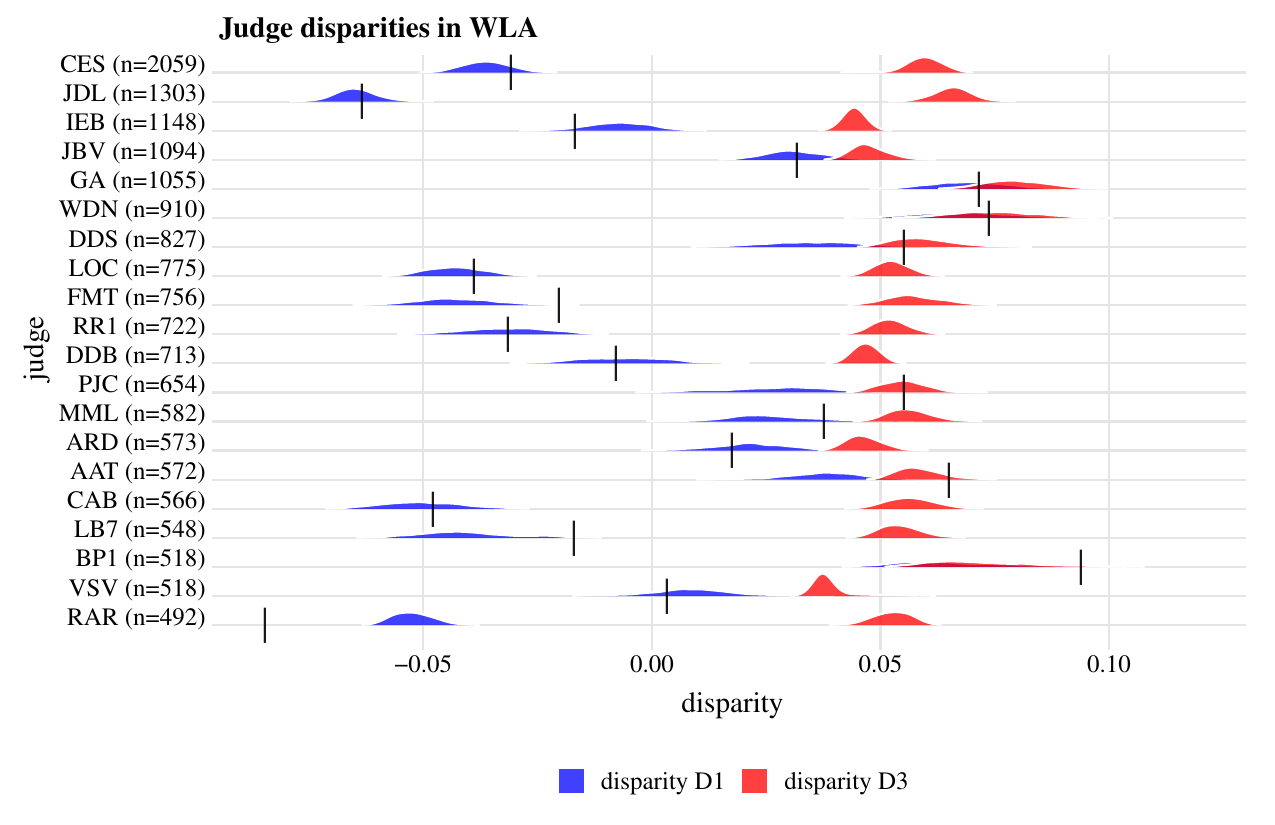}
    \end{subfigure}

\end{figure}

Figure \ref{fig:disparity} displays the posterior distribution of the measures $D_{1}(j)$ and $D_{3}(j)$ for the top 20 (of 119) judges, by caseload, in the New York City court NYC (upper panel) and the top 20 judges (of 98) in the Los Angeles court WLA. 
The difference between judge $j's$ grant rate and the overall grant rate in the court she serves, Equation (\ref{eq:disparity2}), is 
hereafter called ``na\"ive disparity'' and indicated by the black tick mark. Na\"ive disparity is used by RN to infer judicial bias, as 
previously noted. 

In NYC, three judges (AAV, BAN, and SH) grant asylum at least 22\% less often than the court-wide average, and four judges (GTC, NB, PMM, and TAB) grant asylum at least 22\% more often; these judges also have the largest absolute disparities $D_{3}(j)$.  Of these, only BAN appears less disparate after adjusting for case heterogeneity; for the other five, $|D_{1}(j)|$ is almost as large or as large as $D_{3}(j)$, indicating AAV and SH are almost never more likely to grant asylum than their peers,  while PLM, PMM, and TAB are almost never less likely. 
For the others judges, only for JBC
, with na\"ive disparity -5.5\%
, does the 99\% posterior credible interval
of $D_{1}(j)$ cover 0.  
The remaining judges 
are clustered to the right of 0, mostly with mean disparities $D_{1}(j)$ about 3.8\% and $D_{3}(j)$ about 18\%. 

In WLA, the story is quite different.  Whereas the grant rate of 55.8\% in NYC implies the na\"ive judge disparities are upper bounded at this value, in WLA, a judge could deviate from the court-wide rate of 8.8\% by more than 90\%.  But the largest na\"ive disparity is less than $10\%$, and even the  
posterior means of $D_{3}(j)$ are less than $10\%$ for all judges.

\begin{figure}
    \centering
    \caption{Overall disparities $D_3(\cdot)$ by court}
    \label{fig:court_disparity}
            
    \vspace{1em}
   \includegraphics[width=.9\linewidth,
                         trim= 0 1.5cm 0 .71cm, 
                         clip
                         ]{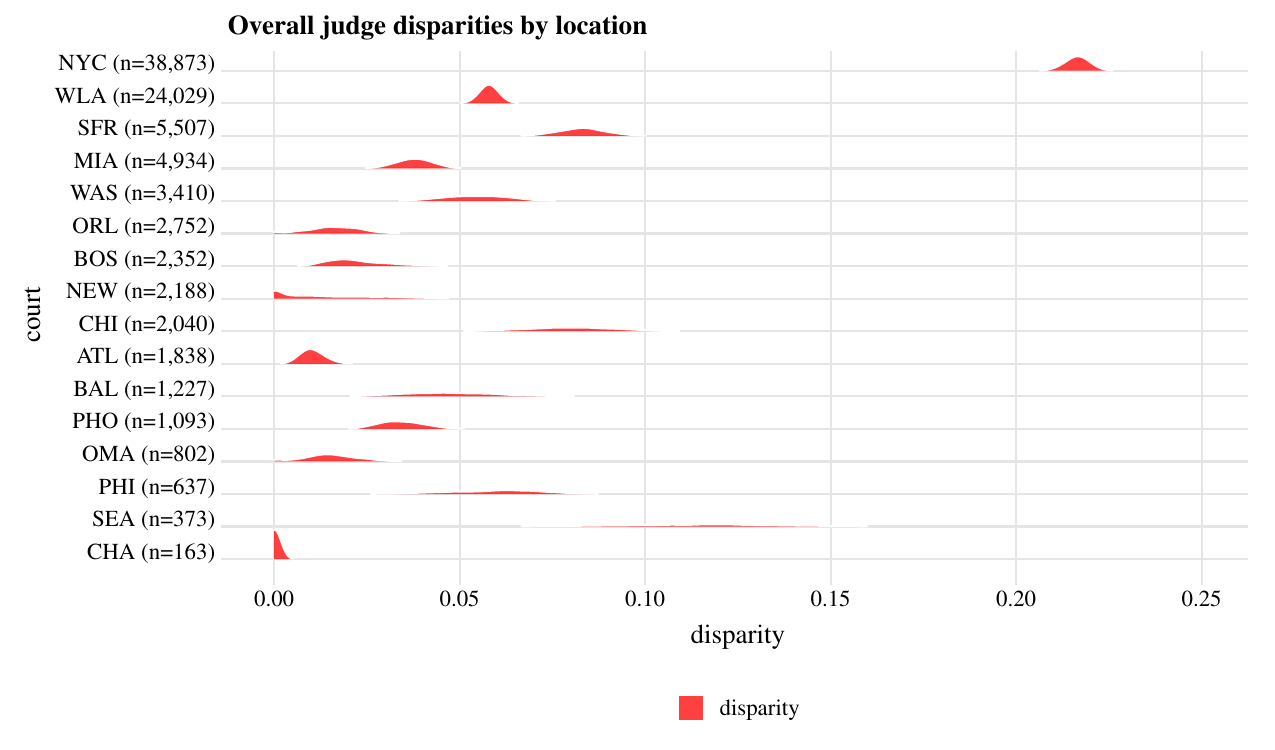}

\end{figure}

Figure \ref{fig:court_disparity} displays the posterior distribution of the absolute court-wide disparities $D_{3}(\cdot)$ for the 16 largest courts. In NYC, with highest grant rate,
the posterior mean of 21\% is nearly four times as large as the mean in WLA of 5.8\%.
But that does not imply a one to one relationship between the court-wide rate and $D_{3}(\cdot)$; for example, in Baltimore (BAL), the posterior mean of 4.8\% is smaller than that in WLA, but the grant rate of 30.8\% is higher.

\begin{figure}
    \centering
    \caption{Judge agreement $A_{1}^{=}(j)$ in New York City (NYC, top) and West Los Angeles (WLA, bottom).}
    \label{fig:agreement}

    \vspace{1em}
    
    \begin{subfigure}{\textwidth}
        \centering
        \caption{NYC}
        \label{fig:agreement_top}
        \vspace{1em}
        \includegraphics[width=.9\linewidth,
                         trim= 0 1.7cm 0 .71cm, 
                         clip
                         ]{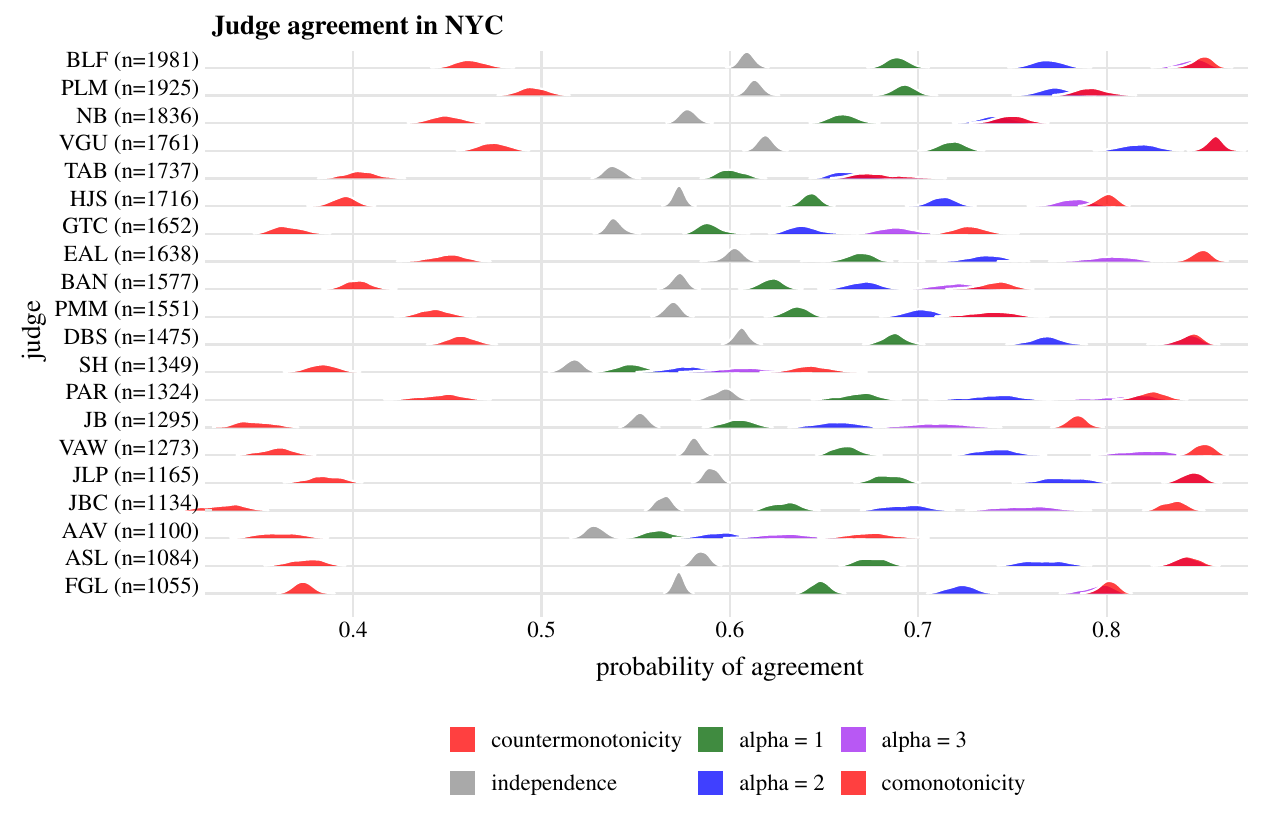}
    \end{subfigure}

    \begin{subfigure}{\textwidth}
        \centering
        \caption{WLA}
        \label{fig:agreement_bottom}
        \vspace{1em}
        \includegraphics[width=.9\linewidth,
                         trim= 0 0 0 .71cm, 
                         clip
                         ]{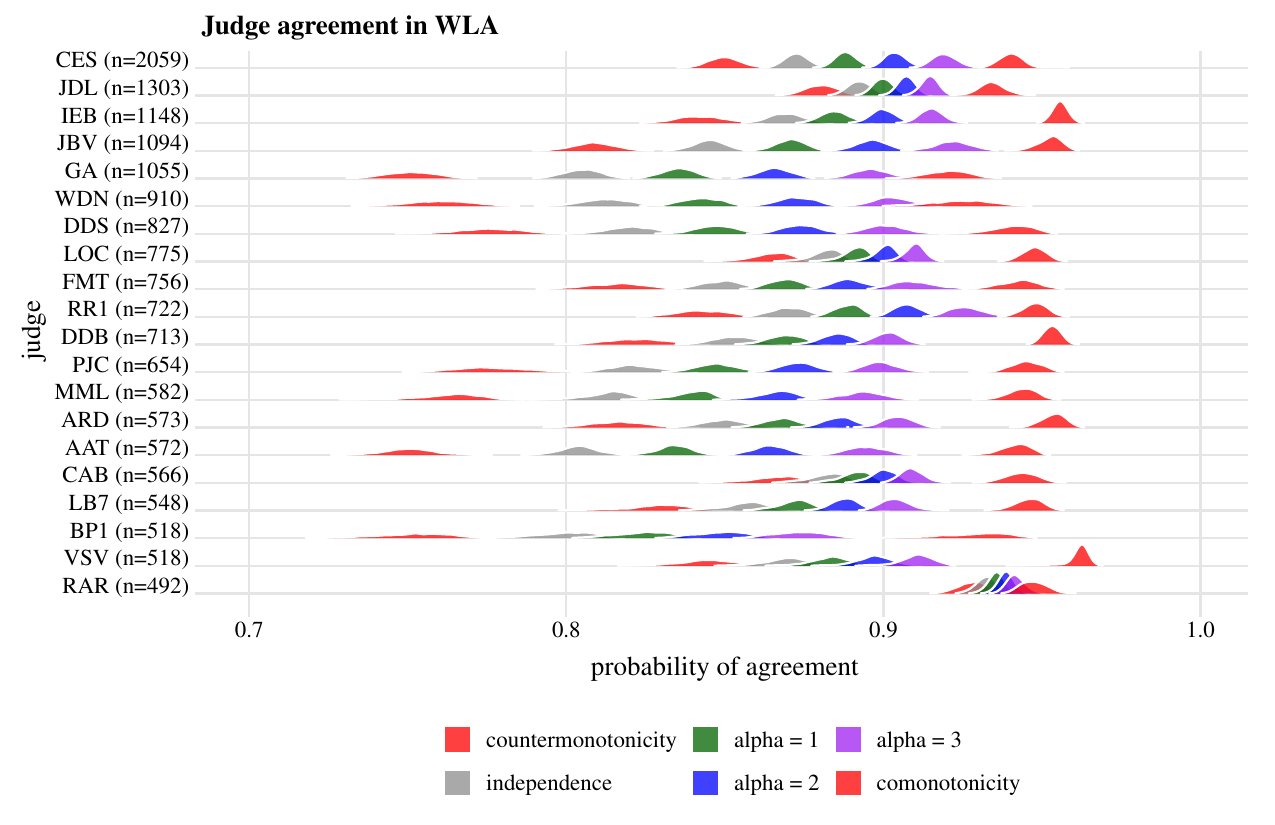}
    \end{subfigure}

\end{figure}

We now turn to agreement.  Figure \ref{fig:agreement} displays the posterior distribution of $A_{1}^{=}(j)$ for the top 20 judges in NYC (top panel) and WLA (bottom panel) under various scenarios. In NYC, the FH lower bound
ranges from about 33\% (JBC) to more than 50\% (PLM).  However, for countermonotonicity to hold,  a judge must weigh the information in $U_{i}$
differently from her peers. We believe it is more reasonable to think judges evaluate such information similarly, as per assumption ({\bf A6}).
If all relevant information is recorded, assumption ({\bf A5}) reduces to $Y_{i}(j) \perp\!\!\!\perp Y_{i}(j^{a}) \mid {\bf V}_{i} = \bf{v}_{i}$, corresponding to independence.  If the unrecorded information ${\bf U}_{i}$ is of equal predictive value (net of the recorded information ${\bf V}_{i}$),
\[\mathbb{E} \left [ \text{Cov}\left (
\mathbb{E}[Y_i(j) \mid {\bf U}_{i}, {\bf V}_{i}], \,
\mathbb{E}[Y_i(j^a) \mid {\bf U}_{i}, {\bf V}_{i}]
\mid {\bf V}_{i} \right ) \right ]
= \text{Cov} \left (
\mathbb{E}[Y_i(j) \mid {\bf V}_{i}], \,
\mathbb{E}[Y_i(j^a) \mid {\bf V}_{i}] \right),\]
corresponding to $\alpha = 1$.  Increasing $\alpha$ further pushes agreement toward comonotonicity, and with $\alpha = 3$, more than half of the judges achieve maximal agreement, with posterior means varying from 61\% for SH to 85.8\% for VGU.

In WLA, agreement is considerably higher.
Under independence, average posterior agreement exceeds 80\%, and agreement between RAR and her peers exceeds 90\%.
Even under countermonotonicity, agreement always exceeds 70\%, and the difference between lower and upper bound agreement never exceeds 20\%.  With such high agreement under independence, there is not much additional relevant information in the unobserved covariates; even with $\alpha = 3$, at best agreement over independence increases by about 9\% for a few judges.

\begin{figure}
    \centering
    \caption{Overall agreement $A_{1}^{=}(\cdot)$ by court}
    \label{fig:court_agreement}
            
    \vspace{1em}
   \includegraphics[width=.9\linewidth,
                         trim= 0 0 0 .71cm, 
                         clip
                         ]{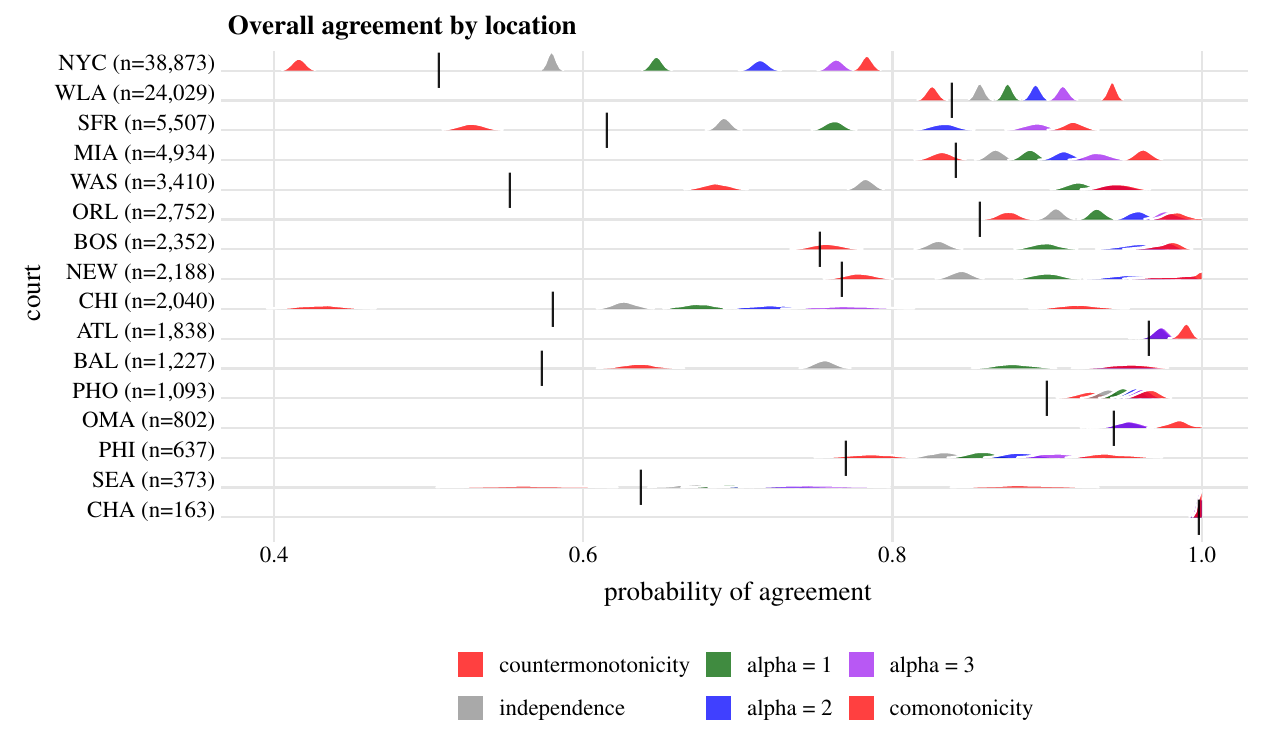}
\end{figure}

Figure \ref{fig:court_agreement} displays court-wide rates of agreement for the top 16 courts. The black tick mark represents agreement under ``na\"ive independence'' $g_{c}^{2} + (1 - g_{c})^{2}$ where $g_{c}$ is the court-wide grant rate.
Courts with the lowest rates  (CHA--0.00\%, ATL--1.72\%, OMA--2.91, PHO--5.28, ORL--7.72\%, MIA--8.68\%) tend to have the most agreement under all scenarios, with lower bounds in excess of 80\%, and agreement under independence in excess of 87\%, with little room for unobserved covariates to push agreement much higher. 
In courts with intermediate rates, e.g., WAS, with overall grant rate of 33.7\%, and BAL with grant rate 30.8\%, adjusting for case heterogeneity and judge heterogeneity matters more.  In WAS, under ``na\"ive independence'' agreement would be 
$55.3\%$.  But even if countermonotonicity were to hold, the posterior mean would exceed 69\%, and the mean under independence (78\%) is almost 42\% higher than under ``na\"ive independence''.  And if $\alpha = 1$, agreement virtually reaches its upper bound of 95\%. But in Chicago, with grant rate 29.9\%, only slightly less than WAS, the posterior mean lower bound of 43\% is roughly 25 percentage points less than in WAS, and under independence, the posterior mean of 63\% is only slightly greater than in NYC, and 15 percentage points less than in WAS.  Further, unlike WAS, choosing $\alpha =1$ only increases agreement by a few (4.8) percentage points relative to independence.  

Finally, we compute the average posterior agreement among judges in the top 16 courts (where 90\% of cases were filed), weighted by the number of cases in each court. 
The FH lower bound is 62.7\%, the FH upper bound, 88.4\%. In between are ``na\"ive independence'' (67.5\%), independence (73\%),  $\alpha = 1$ (78.2)\%,  $\alpha = 2$ (82.7\%), and $\alpha = 3$ (85.9\%).  
Taking $\alpha = 1$, which seems more than reasonable, as the recorded covariates include no information about the individual merits of the case, 
the agreement of 78.2\% is about 16\% higher than under ``na\"ive independence.''

To provide another perspective on the agreement of 78.2\%,  consider the following thought experiment: 8 judges are eligible to hear the case, 7 of whom would grant (or not) asylum were they assigned. If two judges are selected uniformly at random (without replacement), the probability they would agree is 7/8 $\times$ 6/7 = .75, roughly the agreement we find in our analysis with $\alpha = 1$. In fact, even under independence ($\alpha = 0$), we have 73.3\% agreement, consistent with 6 out of 7 judges granting (or not granting) asylum.  Even in NYC, agreement under independence ($\alpha = 0$) is approximately 60\%, consistent with 4 of 5.

\section{Discussion}

Public institutions are often evaluated using a community standard in which the performance of each agent is compared with that of peers. Naive comparisons may fail to account for the varied situations different agents face in the performance of their duties. We define estimands of disparity and agreement that account for this variability and show how to estimate these, at both the agent and institutional levels, when only one agent only is assigned to each case.  

We apply our methodology to the U.S. asylum courts, revisiting earlier work that claimed the outcome of an asylum application depends more on the judge assigned to the case than on case merits. In the New York court, where a third of the cases are handled and grant rates are close to 50\%, the disparities among the 20 judges who handled the most cases are larger than the other courts and agreement is lower. Yet disparity among these 20 judges is limited to six outlier judges with $D_{3}(j)$ greater than .3, and wide-spread disagreement does not characterize the court as a whole. In the other courts, where the grant rates are lower, there is less disparity between judges and high agreement, even though the potential for disparity is greater. 
Our findings suggest that the blanket condemnation of the EOIR judiciary as partial and unprofessional in previous work is largely unsubstantiated.  

To ensure comparability with previous work on the asylum courts, particularly with GAO, we analyzed cases filed between 2007 and 2014. Yet our analysis remains relevant today. First, we challenge the interpretation of statistical evidence that continues to serve as the basis for a growing academic literature, and we provide a methodology for correcting the record using tools from causal inference. Second, the asylum system has undergone considerable stress in recent years due to both external pressures and domestic policies. For example, under Title 42, enacted during the COVID-19-era public health emergency, the United States expelled migrants who crossed the border illegally. Policymakers and advocates are now proposing reforms that consider in part how the institution functioned in the past. Mischaracterizing past performance may lead to misguided reforms.

It is also important 
to update our analysis and examine the more recent workings of the courts. One challenge will be that, due to the dramatic increase in filings following 2014, the average wait times for a hearing now exceeds four years, and many cases are unresolved after six or more years. Thus, any timely analysis must also account for right censoring. 

Comparisons using community standards can also be used to address several issues of current concern that speak to the integrity of the courts, for example, whether judges appointed during the Trump administrations behave differently than judges appointed under the Obama or Biden administrations.  \citet{kim2019empirical} conducted an early investigation of this question, albeit with limitations similar to those discussed in Section 3. More generally, in any institutional setting, to assess whether agents in different subpopulations, e.g., male and female, behave differently, we can use the approach herein, by choosing a female (male) alternate $j^{a}$ when $j$ is male (female) (Sobel et al., 2023). 

Our methodology may also be useful in other settings where a community standard is used to compare agents. For example, large private sector firms often collect ratings from customers on their satisfaction with the agents with whom they interacted. These may then be used to compare agents with one another. In the public sector, teachers are evaluated using value added models and police officers by the number of stops or arrests made. In some instances, two or more agents may jointly handle an interaction with the public, while in others, multiple cases may be simultaneously handled by one agent. In future work, we shall address these cases to estimate the disparities and agreement defined herein.

We also intend to consider new estimands.  Building on the representation of agreement in (\ref{eq:agree_disp_identity}), we will develop new agreement measures applicable to both the discrete and continuous cases with 2 or more agents, and take up their estimation, both when all relevant agents are observed and when they are not. Similarly, we shall develop disparity measures for comparing more than two agents. We shall also extend our analyses to the case of multiple and multidimensional outcomes.  Our analyses excluded mediators, most notably representation:  in future work, we intend to extend the framework to handle mediating variables.

Evaluations using a community standard are becoming more common and more consequential as public institutions become increasingly accountable to the public or politicized (depending on one's perspective). Statisticians have an important role to play in developing methodologies that faithfully portray the integrity of these institutions and the performance of their agents; good measurement protects the integrity of our public institutions, while mismeasurement undermines them, perhaps more than if there were no measurement at all.

\printbibliography[title={References}]
\end{refsection}

\setcounter{section}{0}
\renewcommand{\thesection}{\Alph{section}}
\renewcommand{\thesubsection}{\thesection.\arabic{subsection}}

\newpage

\begin{refsection}

\section*{Supplemental Information}

\section{Data Description}

The Executive Office for Immigration Review (EOIR), a component of the U.S. Department of Justice, maintains administrative case data for immigration court proceedings. EOIR adjudicates immigration cases under authority delegated by the Attorney General, and immigration judges appointed by the Attorney General preside over immigration court hearings. EOIR provides raw case-level data from its case-file electronic database on its website. The results presented here use data accessed on June 1, 2026.

The database is relational with three main tables: the case table, the proceeding table, and the schedule table. The case table contains case-level information such as the nationality of the asylum seeker. The proceeding table records the proceeding or proceedings associated with each case. The schedule table records scheduled hearing sessions associated with proceedings---a proceeding can have multiple scheduled sessions.

A separate court application table records applications filed in immigration court, including asylum applications and their outcomes. Using this table, we identify affirmative asylum applications filed between May 1, 2007 and September 30, 2014. We restrict attention to cases classified as affirmative asylum cases (corresponding to cases in which EOIR starts the ``clock'', i.e., C\_ASY\_TYPE = I), and we retain only the first asylum application associated with the lead case (the primary applicant in cases where multiple family members apply together). Later asylum applications or repeat attempts are excluded. Rider cases are not considered, although the number of riders attached to the lead is retained as a covariate.

For each retained lead case, we identify the first proceeding, the respondent’s first appearance before an immigration judge. We use the immigration judge recorded for that first proceeding as the initially assigned judge. The outcome is the eventual decision on the first asylum application, coded as grant or not grant. Specifically, we consider the outcomes GRANT, FULL GRANT and GRANT WCAT to be a ``grant'', and outcomes ABANDONMENT, DENY, NOT ADJUDICATED, WITHDRAWN, ADMIN CLOSURE, COV/TRANSFER, and OTHER to be a ``not grant.''

\section{Model Sensitivity}

We examine the sensitivity of the findings presented in Section 4.4 to the model and estimation strategy described in Section 4.3. We first discuss the challenges of model misspecification and regularization bias. We then consider three alternative models and show that the estimates from Section 4.4 are largely insensitive to the choice of model. As part of our model sensitivity analysis, we remake Figures 1-4 exactly as described in Section 4.4, except we substitute each alternative model for the estimation of the marginal grant rate. These figures are provided in section D of the Supplement.

\subsection{Model Misspecification and Regularization Bias}

In Section 4.4, we describe how the marginal probabilities were estimated nonparametrically using Bayesian Additive Regression Trees (BART, \cite{chipman2010bart}). An advantage of BART is that it avoids model misspecification bias that can arise in the case of parametric models, where the true model may not be contained within the class of candidate models. Misspecification is particularly problematic when there is little overlap between cases, as discussed by \citet{hill2011bayesian}. 

The disadvantage of nonparametric methods like BART is the potential for regularization bias \citep{li2023bayesian, antonelli23}, arising from the deliberate restriction of the model class to avoid overfitting. Both model misspecification and regularization bias are difficult to assess directly. One solution is to fit a variety of models to see how sensitive the findings are to the choice of model. We take that approach here. We reanalyze the data using the same exact methodology as described in the main text, except we substitute a logistic regression model and a partially general linear model. We refer to these as simply the ``semiparametric model'' and ``parametric model.'' We also compare to another implementation of BART, which we refer to as the ``nonparametric model.''

\subsection{Model Choice}

We consider three alternative models for the probability case $i$ would receive asylum under judge $j$. The first is the alternative semiparametric model $$\Pr\{Y_i(j)=1 \mid \mathbf V_i,\mathbf W_j\} = F\{G(\mathbf V_i) + \mathbf W_j^{\mathsf T} \gamma + b_j\},$$ where $G$ is an unknown prediction function and $F$ is the known link function. The vectors $\mathbf V_i^{b}$ and $\mathbf W_j$ are the case and judge covariates discussed in the main text, and $b_j$ is a random effect for each judge.

We estimate the model using the \texttt{R} package \texttt{stan4bart}. The package samples from the posterior distribution of the function $G$ using Bayesian Additive Regression Trees as implemented in the \texttt{R} package \texttt{bart} and the remaining parameters using the \texttt{R} package \texttt{rstanarm}. We use the default link function (probit), number of trees ($m=200$), and priors (see \citet{gelman2020regression} for details). 

The second is the alternative parametric model (logistic regression), $$\Pr\{Y_i(j)=1 \mid \mathbf V_i,\mathbf W_j\} = F\{\beta_0
  + \mathbf V_i^{\mathsf T}\boldsymbol\beta
  + \mathbf W_j^{\mathsf T}\boldsymbol\gamma
  + b_j
  + a_{n[i]}
  + c_{\ell[i]}
  \}$$ 
where $F$ is the known link function, $\beta_0$ is an intercept, and $b_{j}$, $n[i]$ and $\ell[i]$ are random effects for judge, nationality, and language, respectively. We sample from the posterior distribution of the model parameters using the \texttt{R} package \texttt{brms} and default priors, see \citet{gelman2020regression} for details. 

We also compare these models to the alternative nonparametric model described in the main text, fit using the \texttt{R} package \texttt{BART} and the default logit link function. For all models, we draw 7,000 samples from the posterior distribution. We discard the first 2,000 as warm-up, yielding 5,000 draws for inference as described in the main text. To ensure that the posterior draws are not sensitive to the choice of seed, we check for posterior convergence. Specifically, we run each model an additional four times (i.e. four chains with different seeds) and observe that the chains have mixed as indicated by a R-hat close to 1 and provide the same estimates as the original draws. Since the estimates are the same, we do not use these additional draws.

\section{Additional Tables}
\label{sec:appendix_tables}
\setstretch{1}
\begin{table}[ht]
\centering
\caption{Case counts, judge counts, grant rates, and agreement under na\"ive independence (na\"ive agreement) by location}
\label{tab:summary}
\small
\begin{tabular}{lcccccc}
\toprule
\makecell{Court\\Location} & \makecell{Number of\\Cases} & \makecell{Number of\\Judges} & \makecell{Grant Rate\\(\%)} & \makecell{Number of\\Cases with\\Alternatives} & \makecell{Number of\\Judges with\\Alternatives} & \makecell{Na\"ive Agreement\\(\%)} \\
\midrule
NYC & 39072 & 119 & 55.8 & 38873 & 119 & 50.7\\

WLA & 24546 & 98 & 8.8 & 24029 & 98 & 83.9\\

SFR & 5856 & 77 & 26.0 & 5507 & 76 & 61.5\\

MIA & 5347 & 52 & 8.7 & 4934 & 52 & 84.2\\

WAS & 4554 & 53 & 33.7 & 3410 & 53 & 55.3\\

ORL & 3667 & 20 & 7.7 & 2752 & 20 & 85.7\\

BOS & 2926 & 40 & 14.4 & 2352 & 38 & 75.3\\

NEW & 2798 & 36 & 13.4 & 2188 & 36 & 76.8\\

CHI & 2565 & 37 & 29.9 & 2040 & 37 & 58.1\\

BAL & 2279 & 31 & 30.8 & 1227 & 28 & 57.4\\

ATL & 2206 & 26 & 1.7 & 1838 & 25 & 96.6\\

PHO & 1912 & 25 & 5.3 & 1093 & 24 & 90.0\\

CHA & 1329 & 6 & 0.1 & 163 & 6 & 99.8\\

PHI & 1268 & 17 & 13.2 & 637 & 17 & 77.0\\

OMA & 1167 & 11 & 2.9 & 802 & 9 & 94.3\\

SEA & 896 & 15 & 23.8 & 373 & 15 & 63.8\\
\bottomrule
\end{tabular}
\end{table}

\vspace{2em}

\section{Additional Figures}

\begin{figure}[ht]
    \centering
    \caption{Judge disparities $D_1(j)$ and $D_{3}(j)$ in New York City using the alternative nonparametric model (top), semiparametric model (middle), and parametric model (bottom).}
    \label{fig:disparity_sensitivity_nyc}


    \begin{subfigure}{\textwidth}
        \centering
        \caption{Nonparametric}
        \label{fig:disparity_top_bart}
        \includegraphics[width=.65\linewidth,
                         trim= 0 1cm 0 .76cm, 
                         clip
                         ]{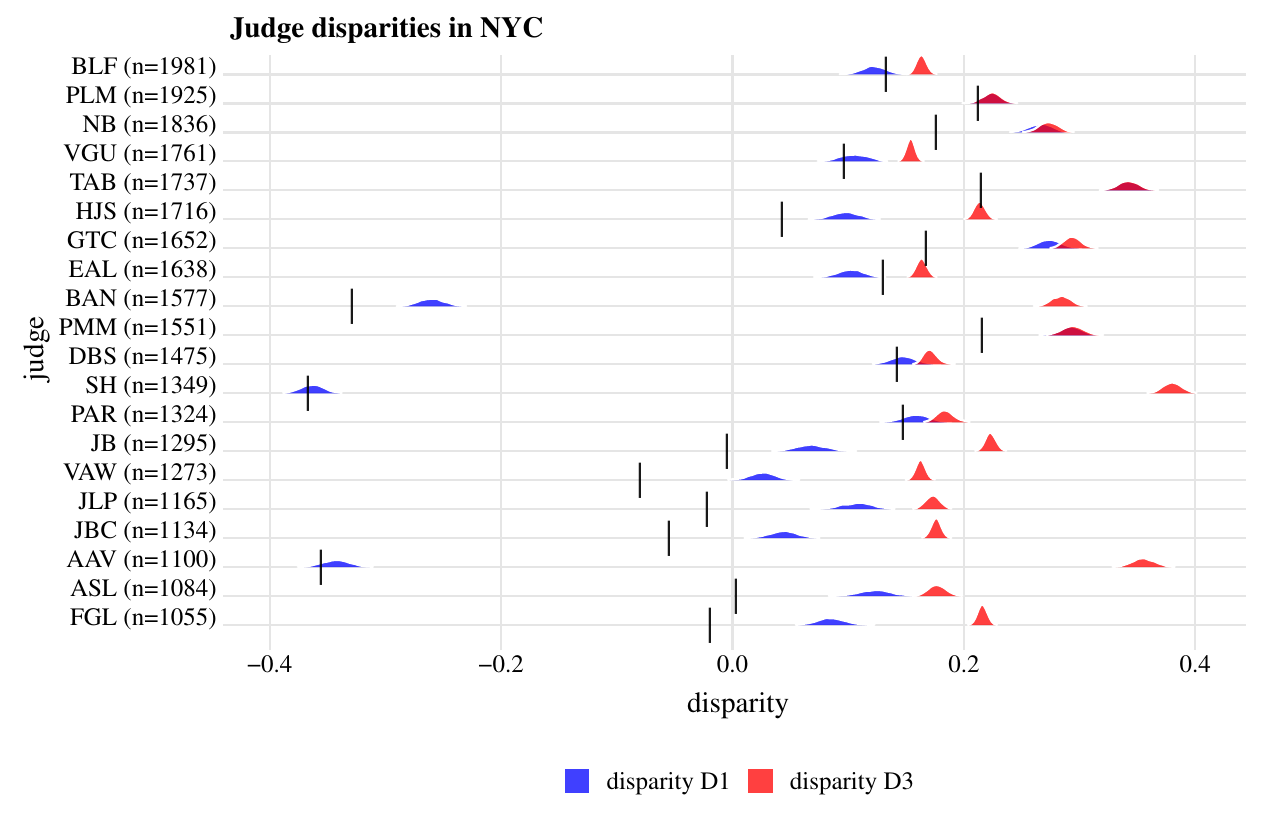}

    \end{subfigure}
        
    \begin{subfigure}{\textwidth}
        \centering
        \caption{Semiparametric}
        \label{fig:disparity_top_stan4bart}
        \includegraphics[width=.65\linewidth,
                         trim= 0 1cm 0 .76cm, 
                         clip
                         ]{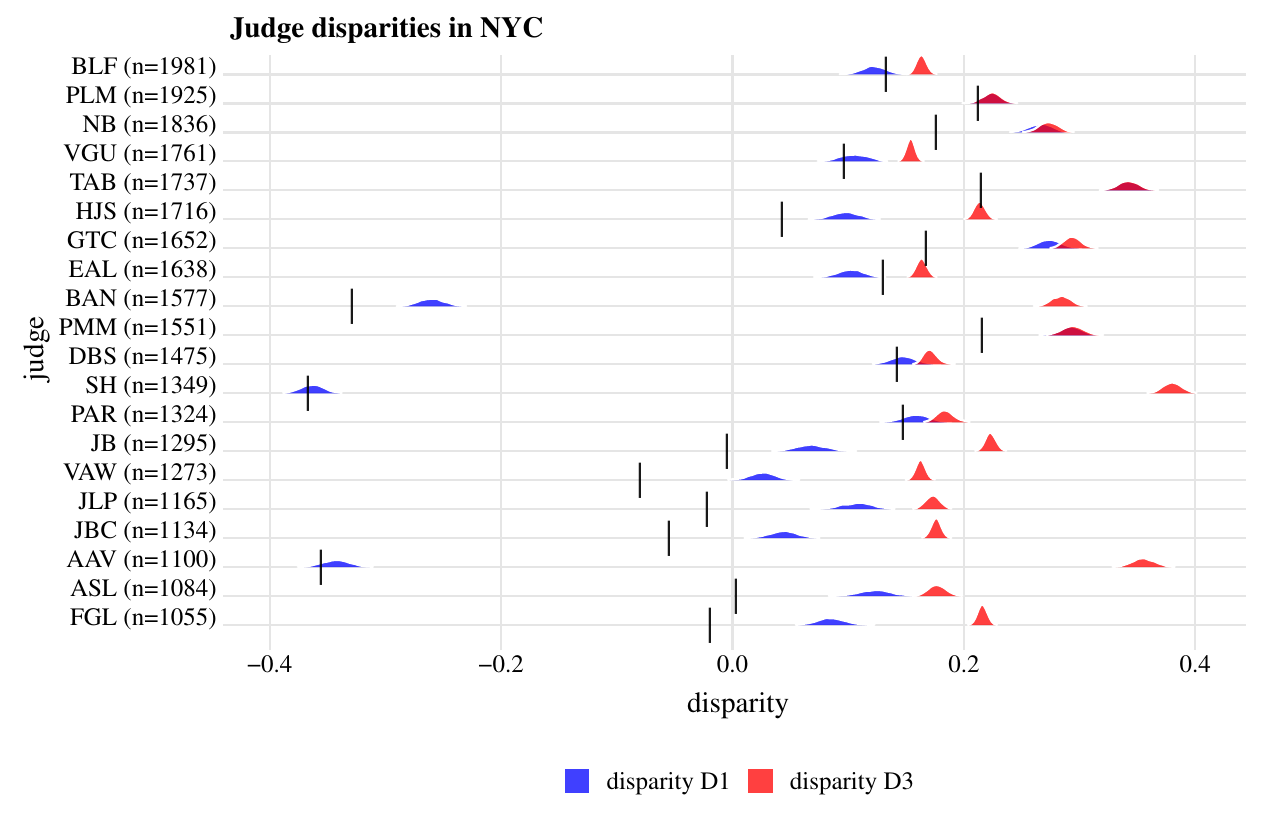}

    \end{subfigure}

    \begin{subfigure}{\textwidth}
        \centering
        \caption{Parametric}
        \label{fig:disparity_top_brms}
        \includegraphics[width=.65\linewidth,
                         trim= 0 0 0 .76cm, 
                         clip
                         ]{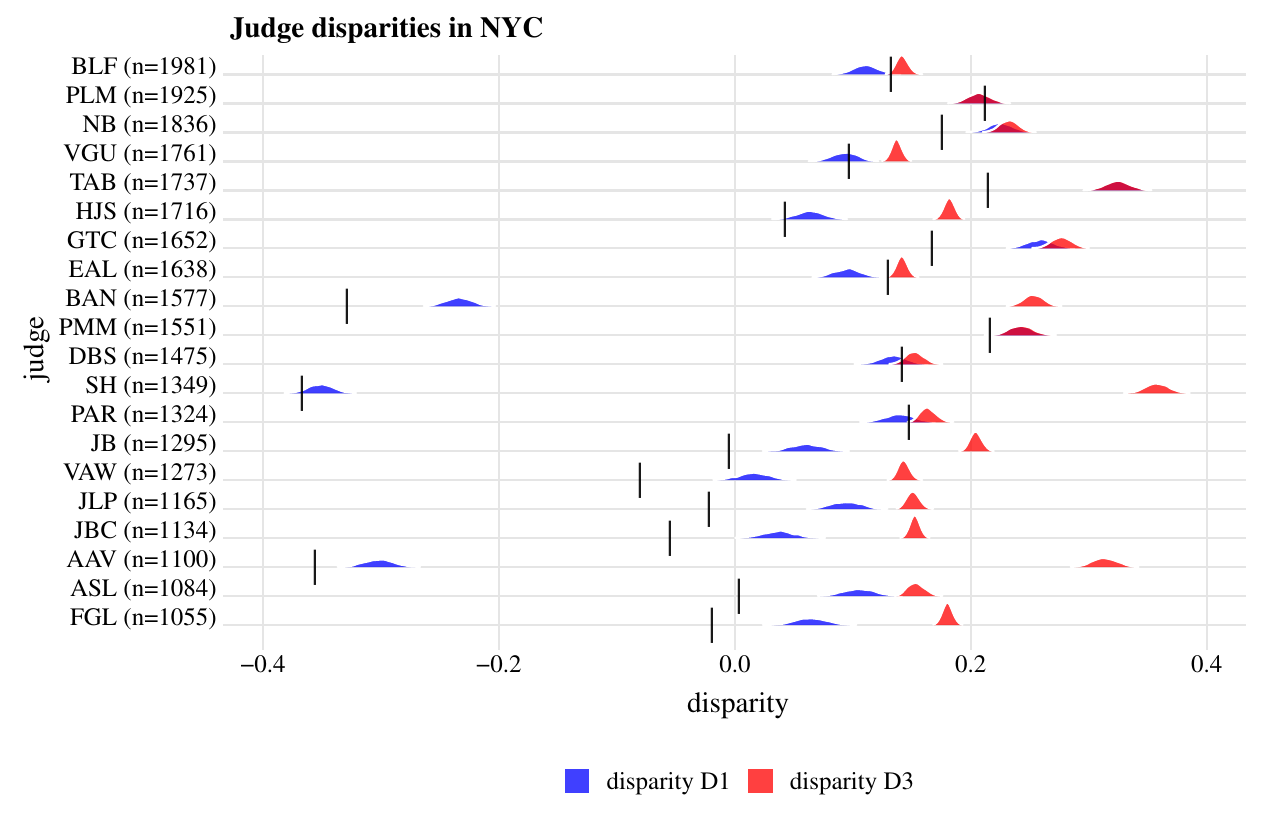}
    \end{subfigure}

\end{figure}    

\begin{figure}[ht]
    \centering
    \caption{Judge disparities $D_1(j)$ and $D_{3}(j)$ in West Los Angeles using the alternative nonparametric model (top), semiparametric model (middle), and parametric model (bottom).}
    \label{fig:disparity_sensitivity_wla}


    \begin{subfigure}{\textwidth}
        \centering
        \caption{Nonparametric}
        \label{fig:disparity_bottom_bart}
        \includegraphics[width=.65\linewidth,
                         trim= 0 1cm 0 .76cm, 
                         clip
                         ]{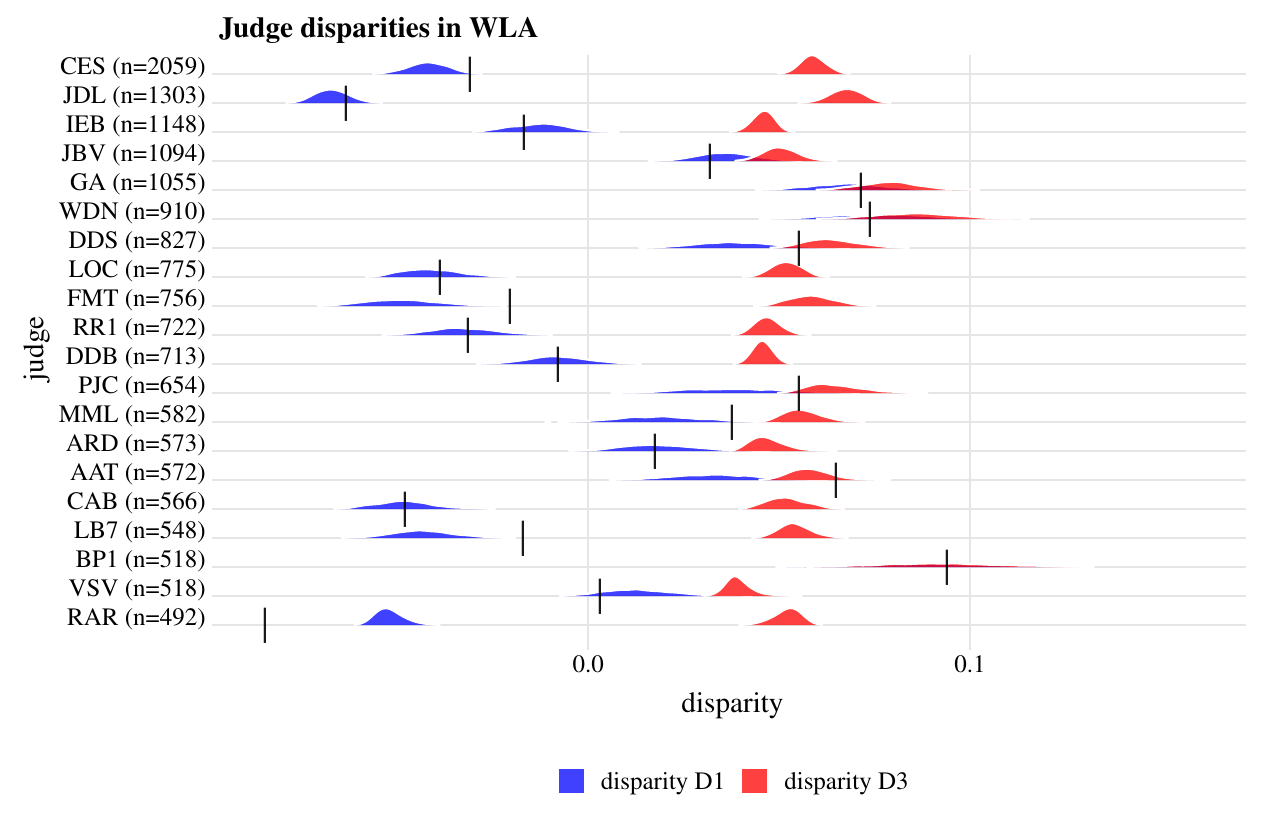}

    \end{subfigure}
        
    \begin{subfigure}{\textwidth}
        \centering
        \caption{Semiparametric}
        \label{fig:disparity_bottom_stan4bart}
        \includegraphics[width=.65\linewidth,
                         trim= 0 1cm 0 .76cm, 
                         clip
                         ]{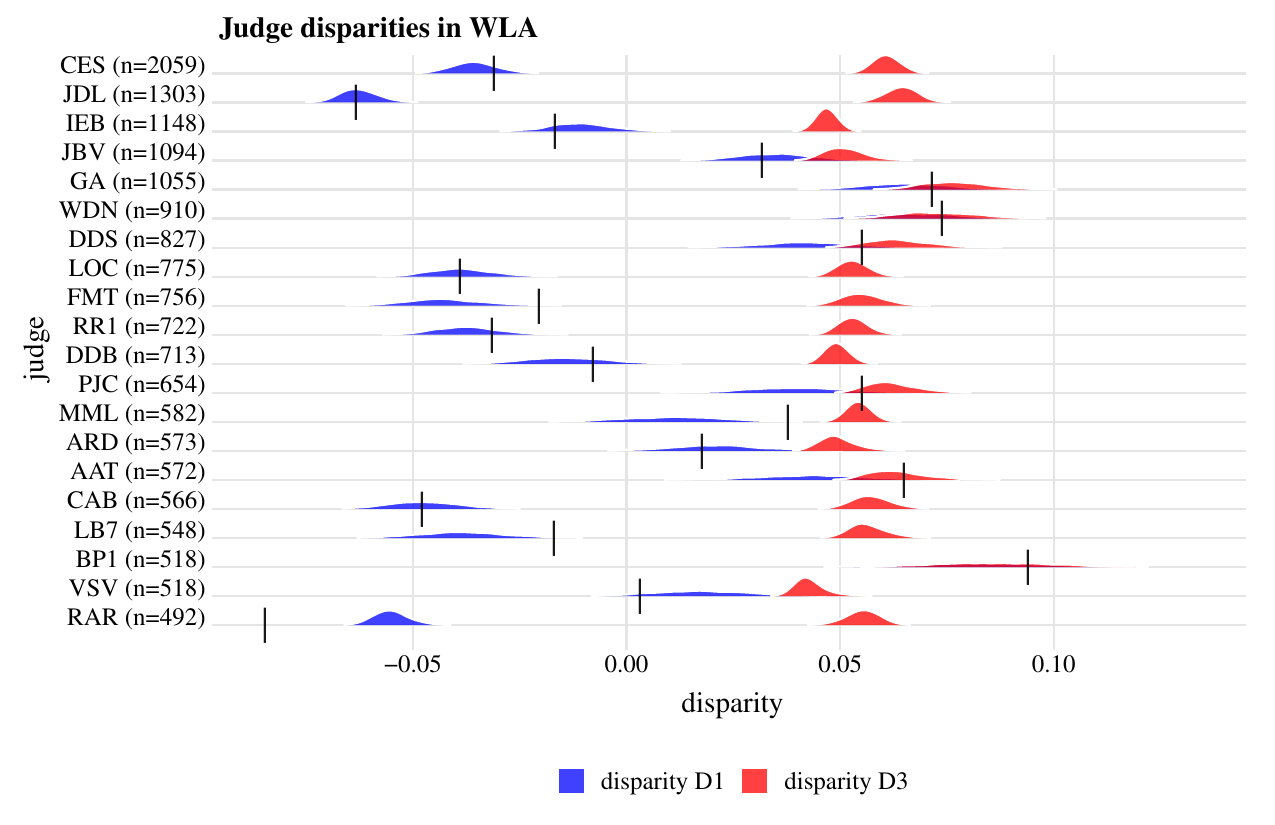}

    \end{subfigure}

    \begin{subfigure}{\textwidth}
        \centering
        \caption{Parametric}
        \label{fig:disparity_bottom_brms}
        \includegraphics[width=.65\linewidth,
                         trim= 0 0 0 .76cm, 
                         clip
                         ]{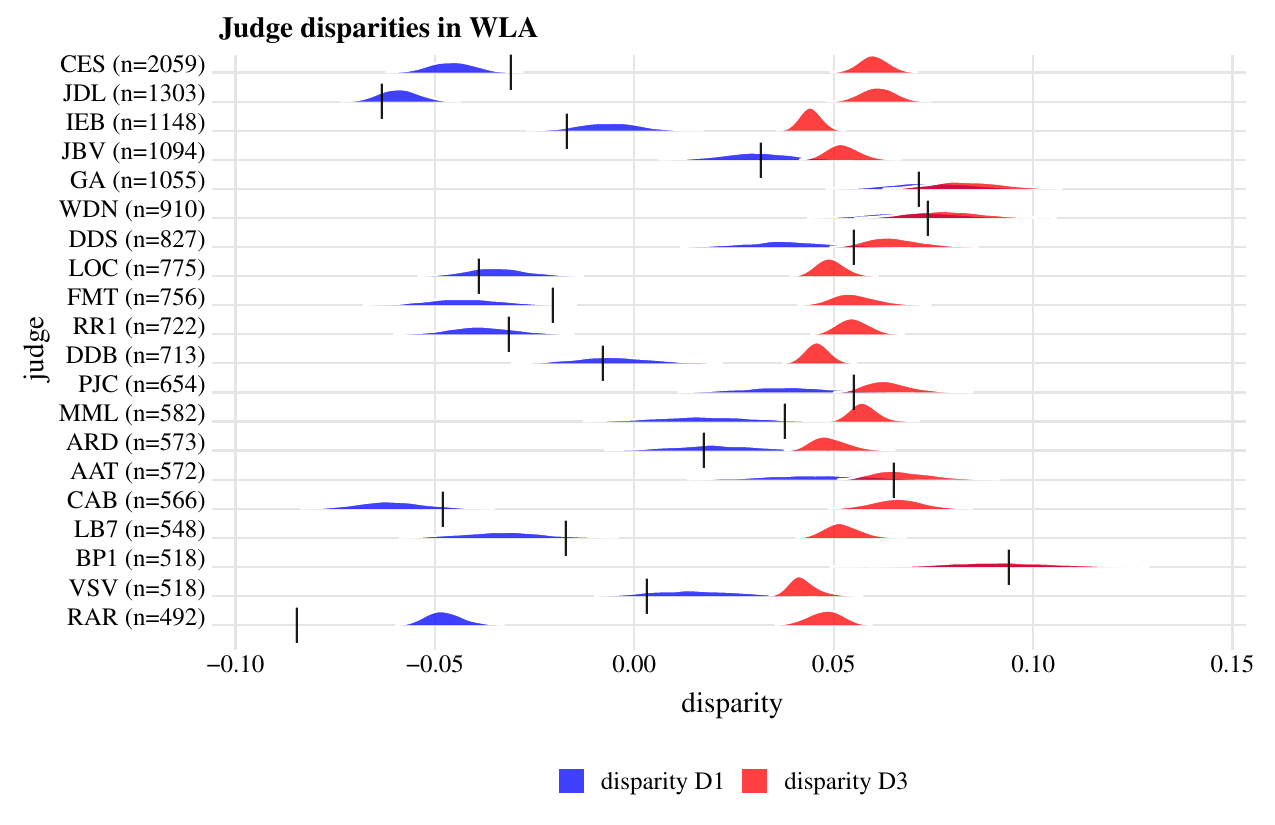}
    \end{subfigure}

    
\end{figure}    

\begin{figure}[ht]
    \centering
    \caption{Overall disparities $D_3(\cdot)$ by court using the nonparametric model (top), semiparametric model (middle), and parametric model (bottom).}
    \label{fig:court_disparity_sensitivity}


    \begin{subfigure}{\textwidth}
        \centering
        \caption{Nonparametric}
        \label{fig:location_disparity_bottom_bart}
        \includegraphics[width=.65\linewidth,
                         trim= 0 1cm 0 .76cm, 
                         clip
                         ]{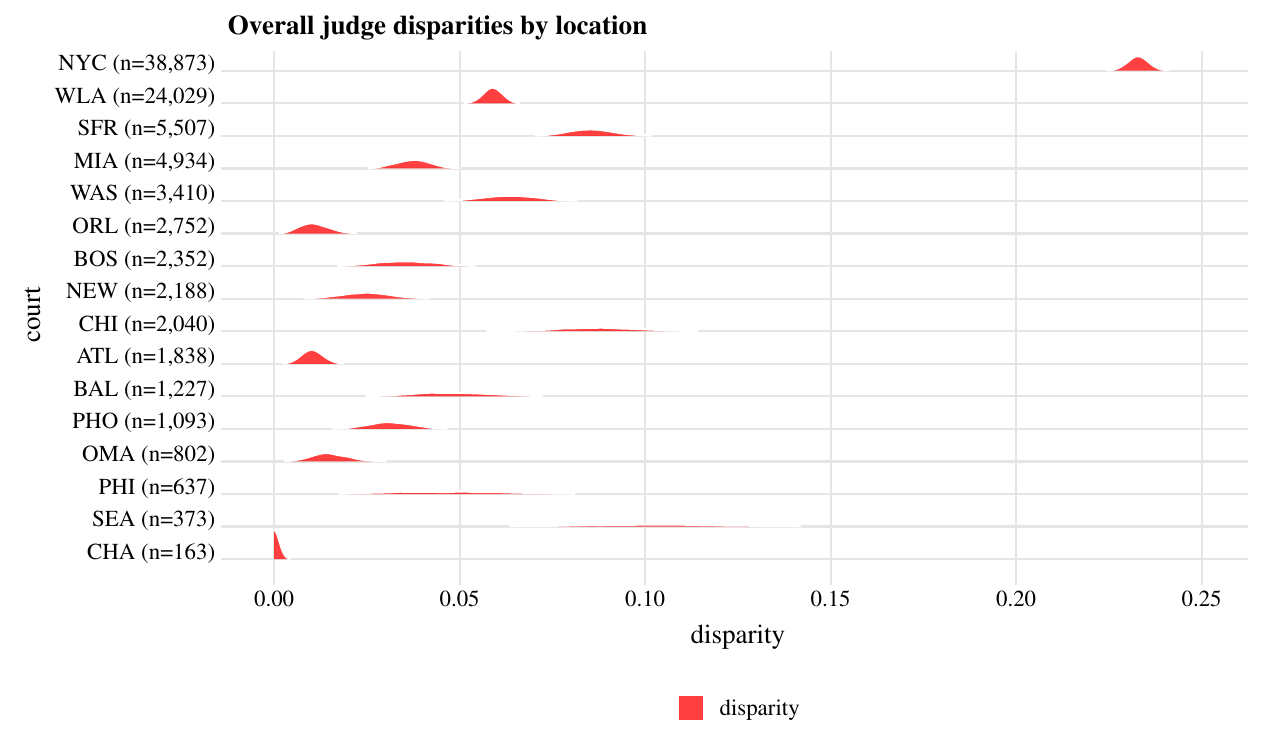}

    \end{subfigure}

    \begin{subfigure}{\textwidth}
        \centering
        \caption{Semiparametric}
        \label{fig:location_disparity_bottom_stan4bart}
        \includegraphics[width=.65\linewidth,
                         trim= 0 1cm 0 .76cm, 
                         clip
                         ]{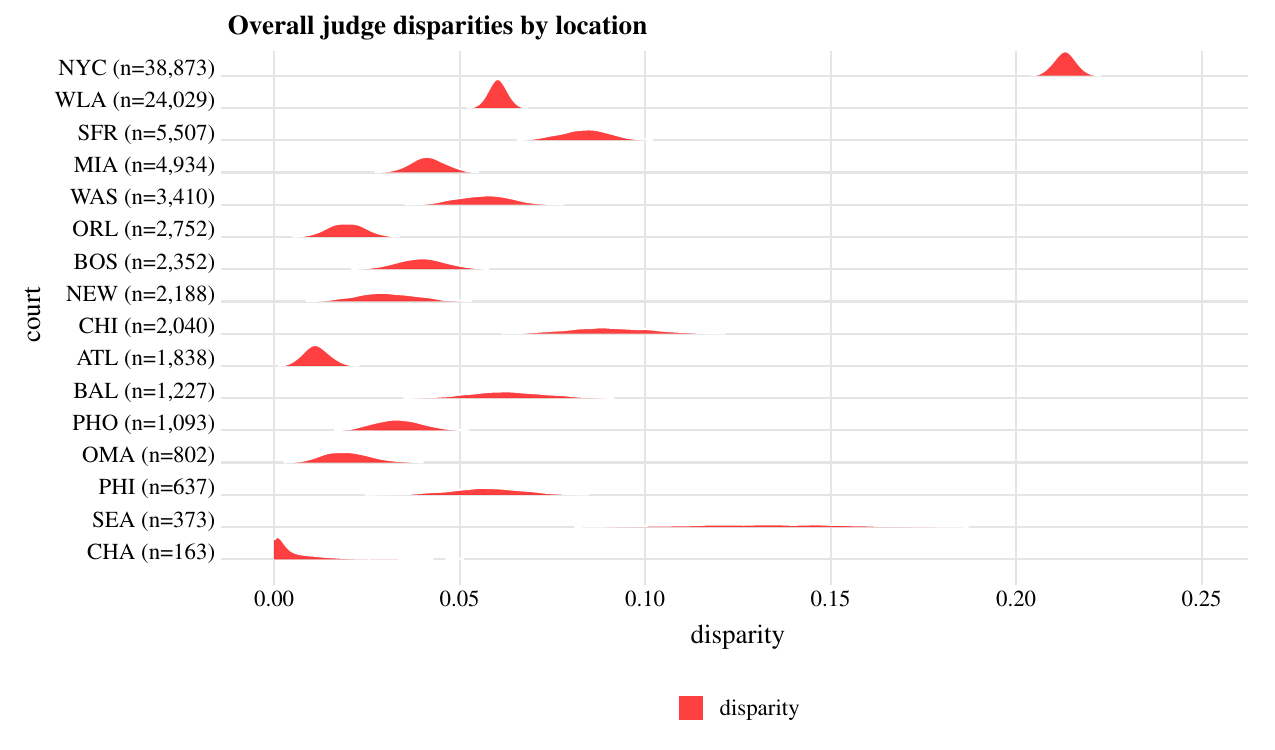}

    \end{subfigure}

    \begin{subfigure}{\textwidth}
        \centering
        \caption{Parametric}
        \label{fig:location_disparity_bottom_brms}
        \includegraphics[width=.65\linewidth,
                         trim= 0 0 0 .76cm, 
                         clip
                         ]{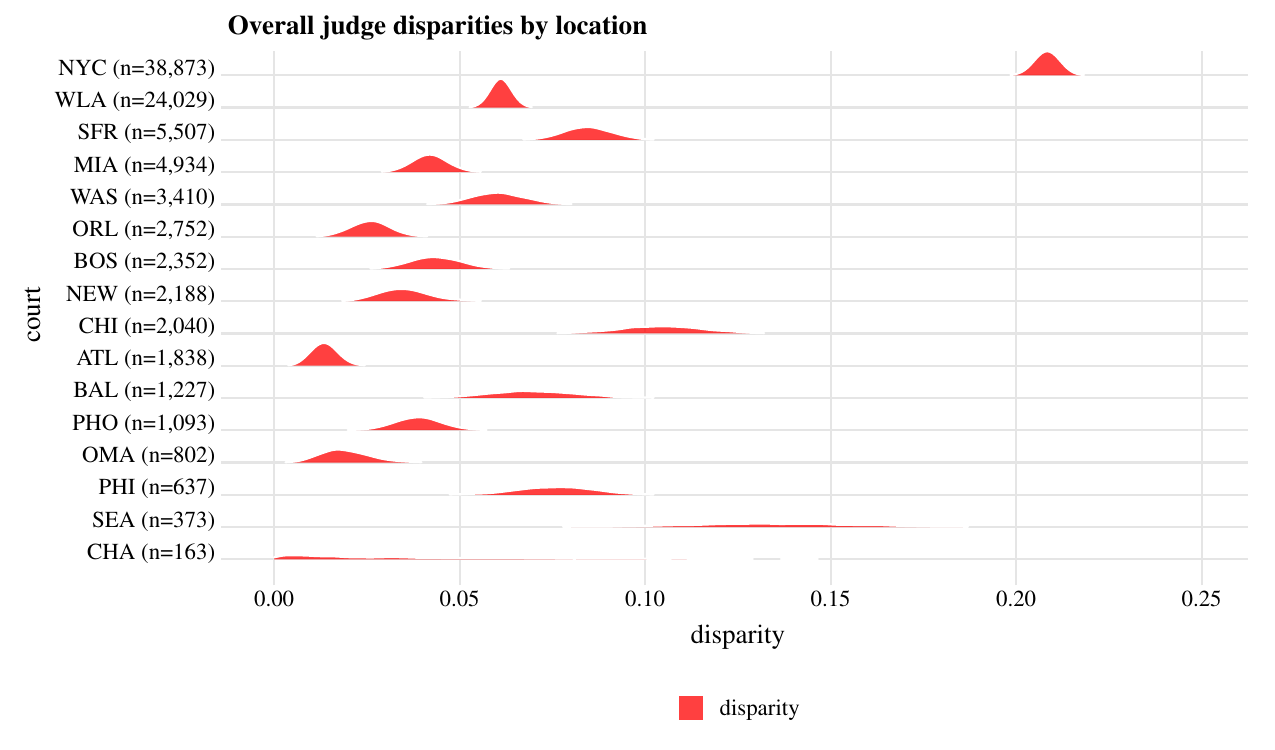}
    \end{subfigure}    
\end{figure}

\begin{figure}[ht]
    \centering
    \caption{Judge agreement $A_{1}^{=}(j)$ in New York City using the alternative nonparametric model (top), semiparametric model (middle), and parametric model (bottom).}
    \label{fig:agreement_sensitivity_nyc}


    \begin{subfigure}{\textwidth}
        \centering
        \caption{Nonparametric}
        \label{fig:nyc_agreement_top_bart}
        \includegraphics[width=.65\linewidth,
                         trim= 0 2cm 0 .76cm, 
                         clip
                         ]{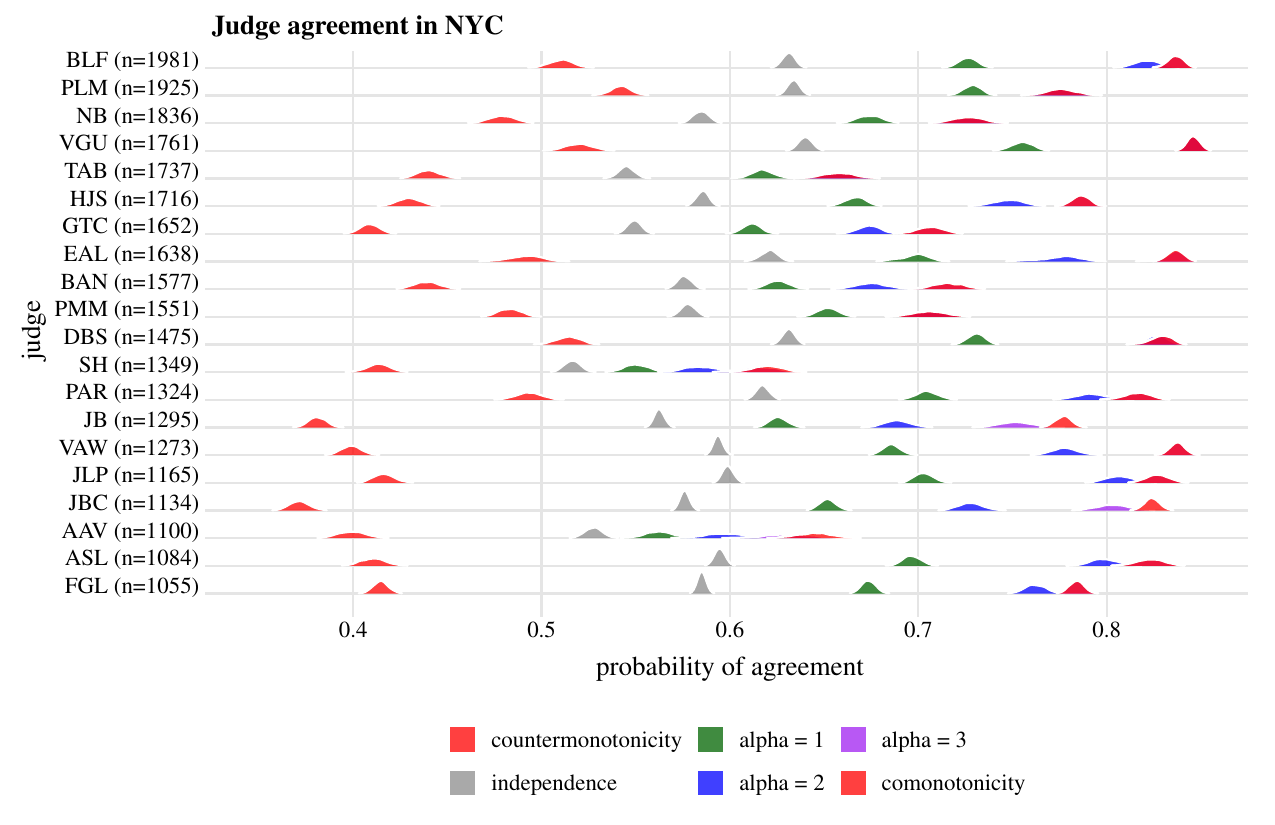}

    \end{subfigure}
        
    \begin{subfigure}{\textwidth}
        \centering
        \caption{Semiparametric}
        \label{fig:nyc_agreement_top_stan4bart}
        \includegraphics[width=.65\linewidth,
                         trim= 0 2cm 0 .76cm, 
                         clip
                         ]{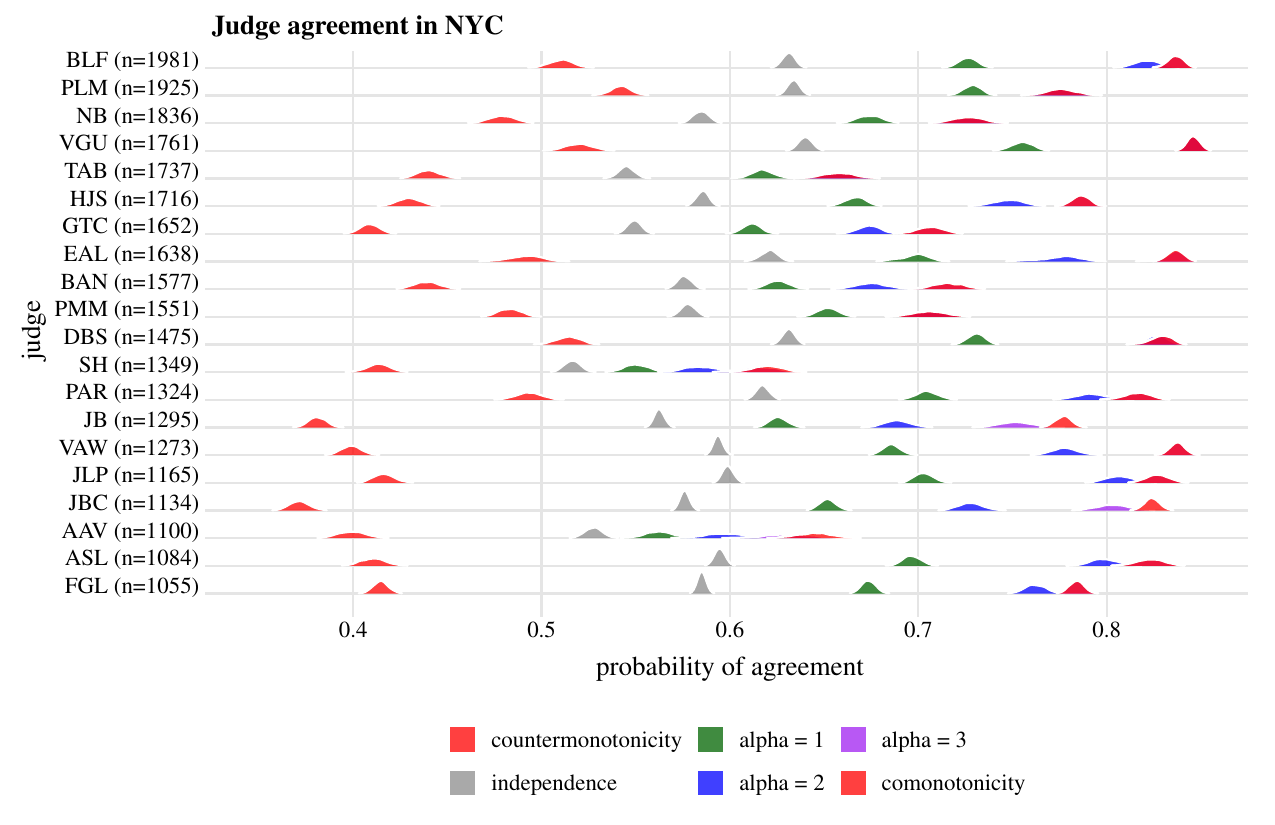}

    \end{subfigure}

    \begin{subfigure}{\textwidth}
        \centering
        \caption{Parametric}
        \label{fig:agreement_top_brms}
        \includegraphics[width=.65\linewidth,
                         trim= 0 0 0 .76cm, 
                         clip
                         ]{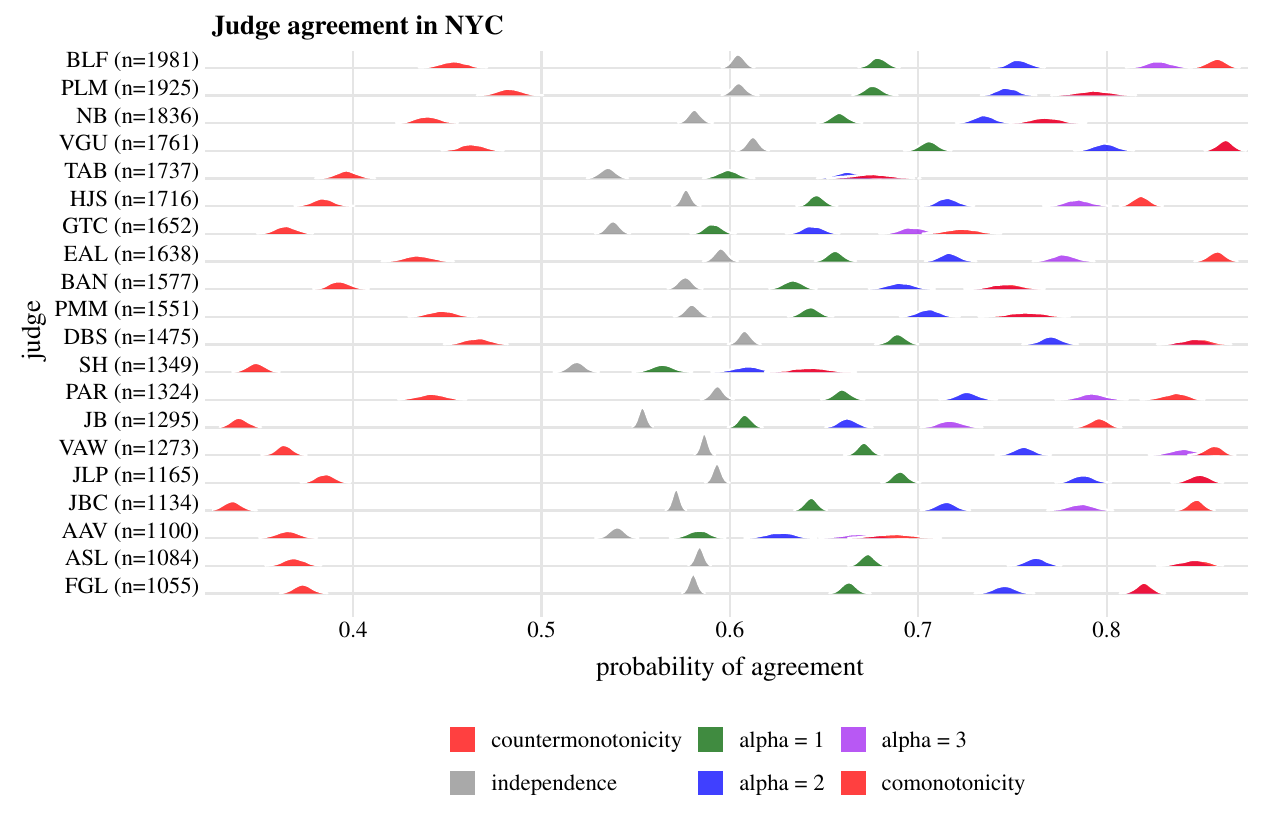}
    \end{subfigure}
\end{figure}    

\begin{figure}[ht]
    \centering
    \caption{Judge agreement $A_{1}^{=}(j)$ in West Los Angeles using the alternative nonparametric model (top), semiparametric model (middle), and parametric model (bottom).}
    \label{fig:agreement_sensitivity_wla}


    \begin{subfigure}{\textwidth}
        \centering
        \caption{Nonparametric}
        \label{fig:wla_agreement_top_bart}
        \includegraphics[width=.65\linewidth,
                         trim= 0 2cm 0 .76cm, 
                         clip
                         ]{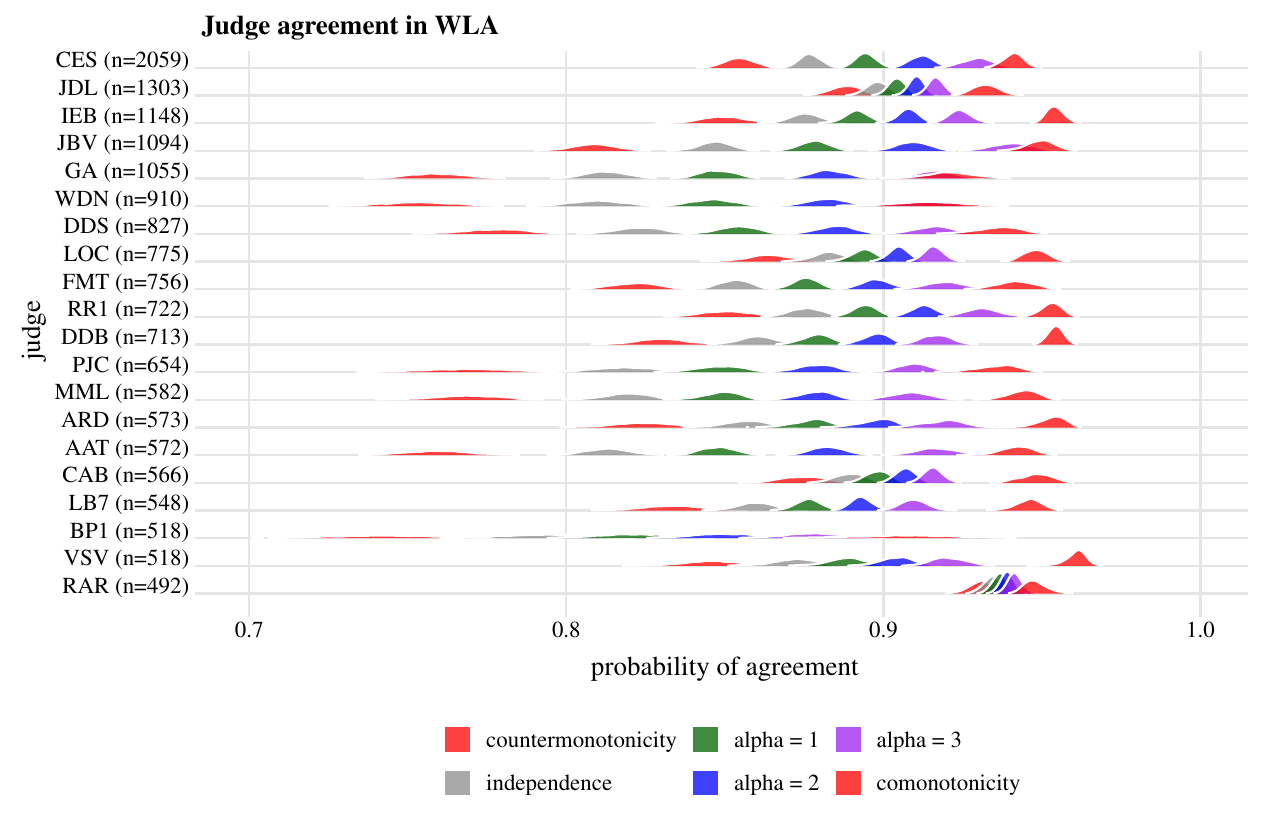}

    \end{subfigure}
        
    \begin{subfigure}{\textwidth}
        \centering
        \caption{Semiparametric}
        \label{fig:wla_agreement_top_stan4bart}
        \includegraphics[width=.65\linewidth,
                         trim= 0 2cm 0 .76cm, 
                         clip
                         ]{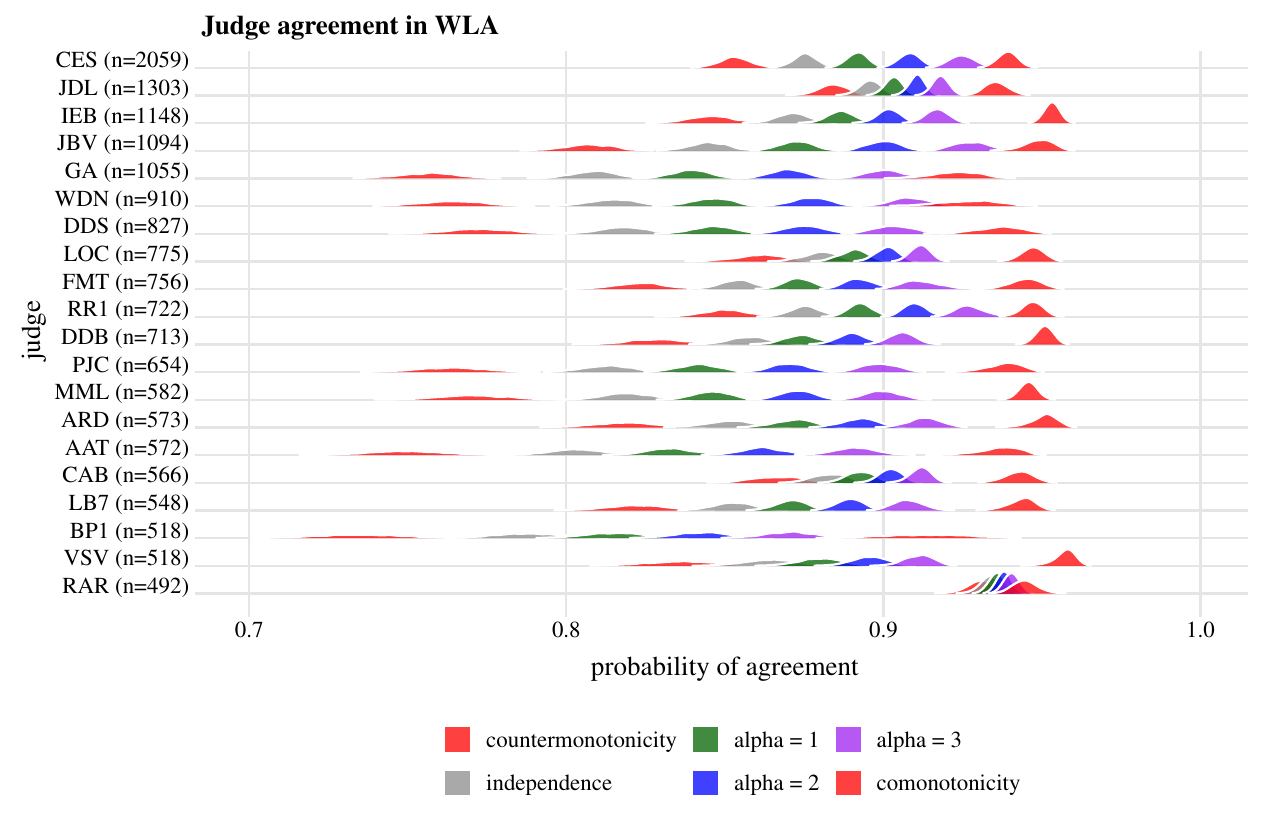}

    \end{subfigure}

    \begin{subfigure}{\textwidth}
        \centering
        \caption{Parametric}
        \label{fig:wla_agreement_top_brms}
        \includegraphics[width=.65\linewidth,
                         trim= 0 0 0 .76cm, 
                         clip
                         ]{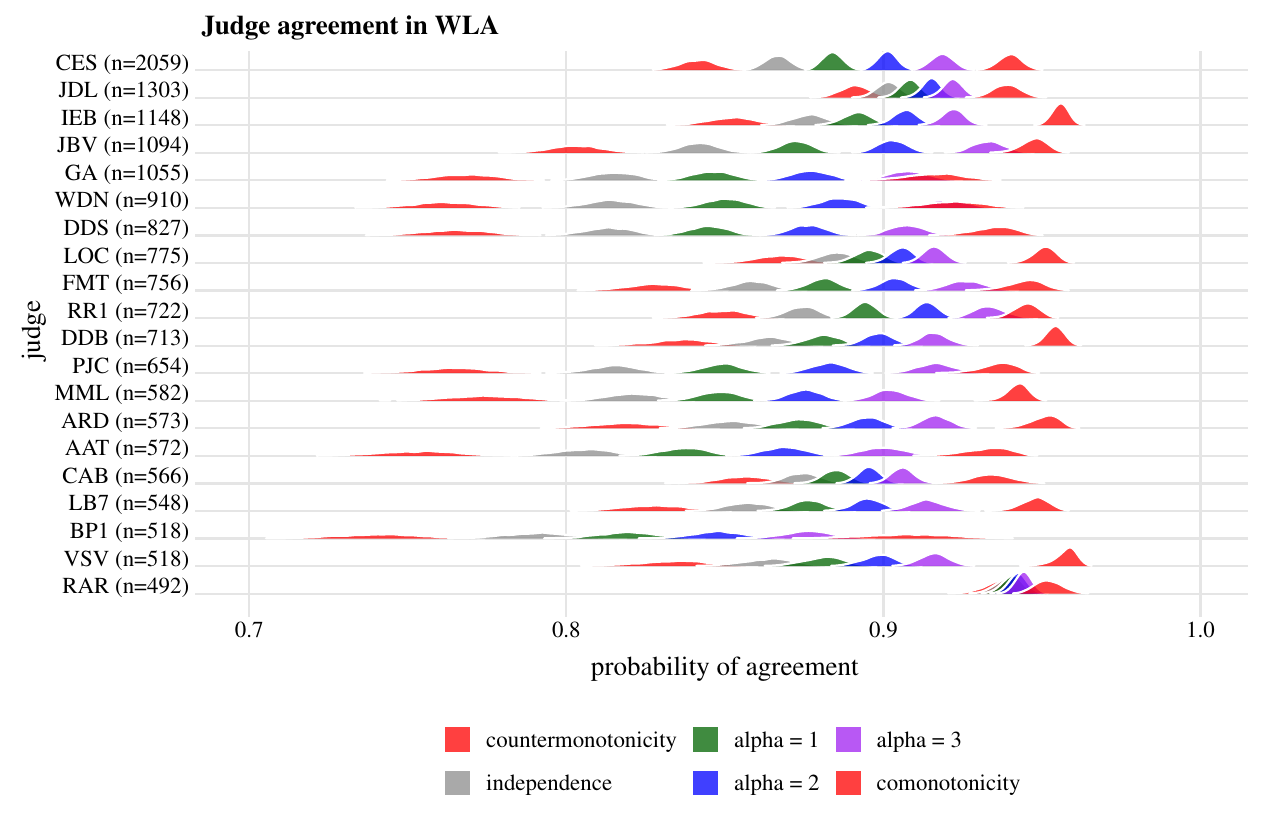}
    \end{subfigure}

\end{figure}    

\begin{figure}[ht]
    \centering
    \caption{Overall agreement $A_{1}^{=}(\cdot)$ by court using the alternative nonparametric model (top), semiparametric model (middle), and parametric model (bottom).}
    \label{fig:location_agreement_sensitivity_nyc}


    \begin{subfigure}{\textwidth}
        \centering
        \caption{Nonparametric}
        \label{fig:location_agreement_bart}
        \vspace{1em}
        \includegraphics[width=.65\linewidth,
                         trim= 0 2cm 0 .76cm, 
                         clip
                         ]{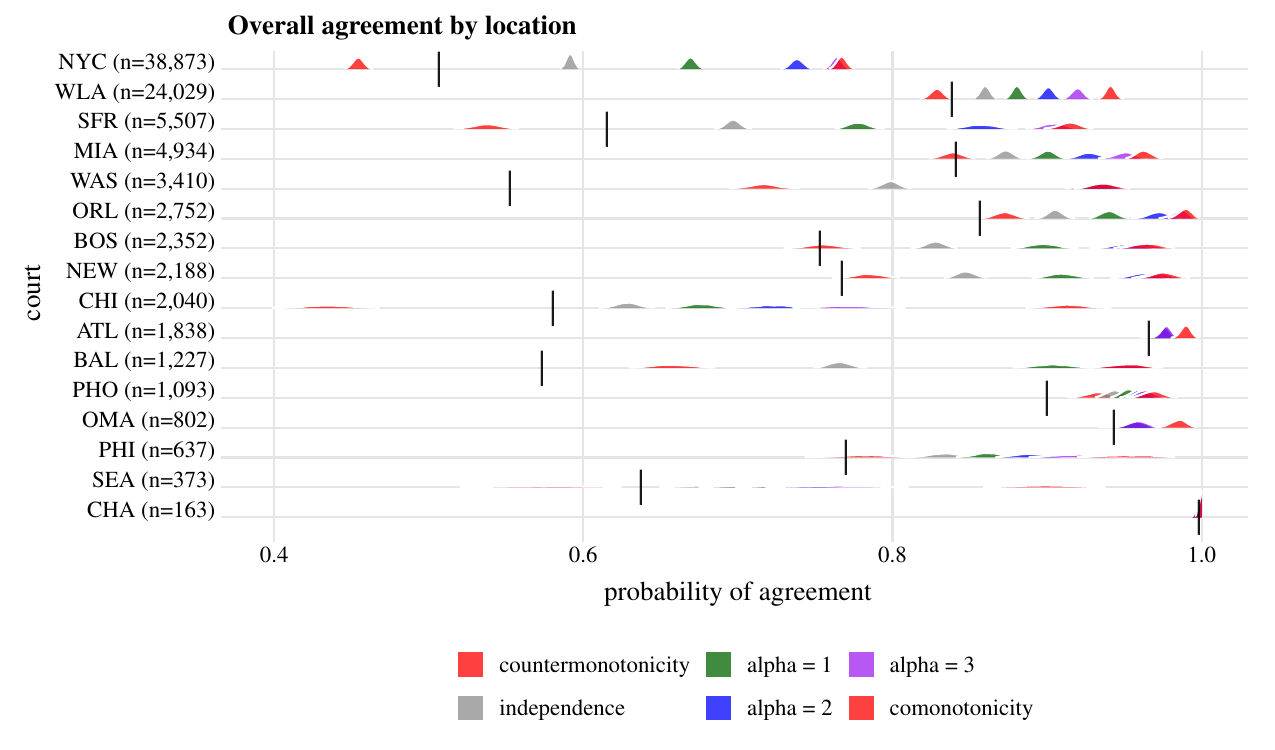}

    \end{subfigure}
        
    \begin{subfigure}{\textwidth}
        \centering
        \caption{Semiparametric}
        \label{fig:location_agreement_stan4bart}
        \vspace{1em}
        \includegraphics[width=.65\linewidth,
                         trim= 0 2cm 0 .76cm, 
                         clip
                         ]{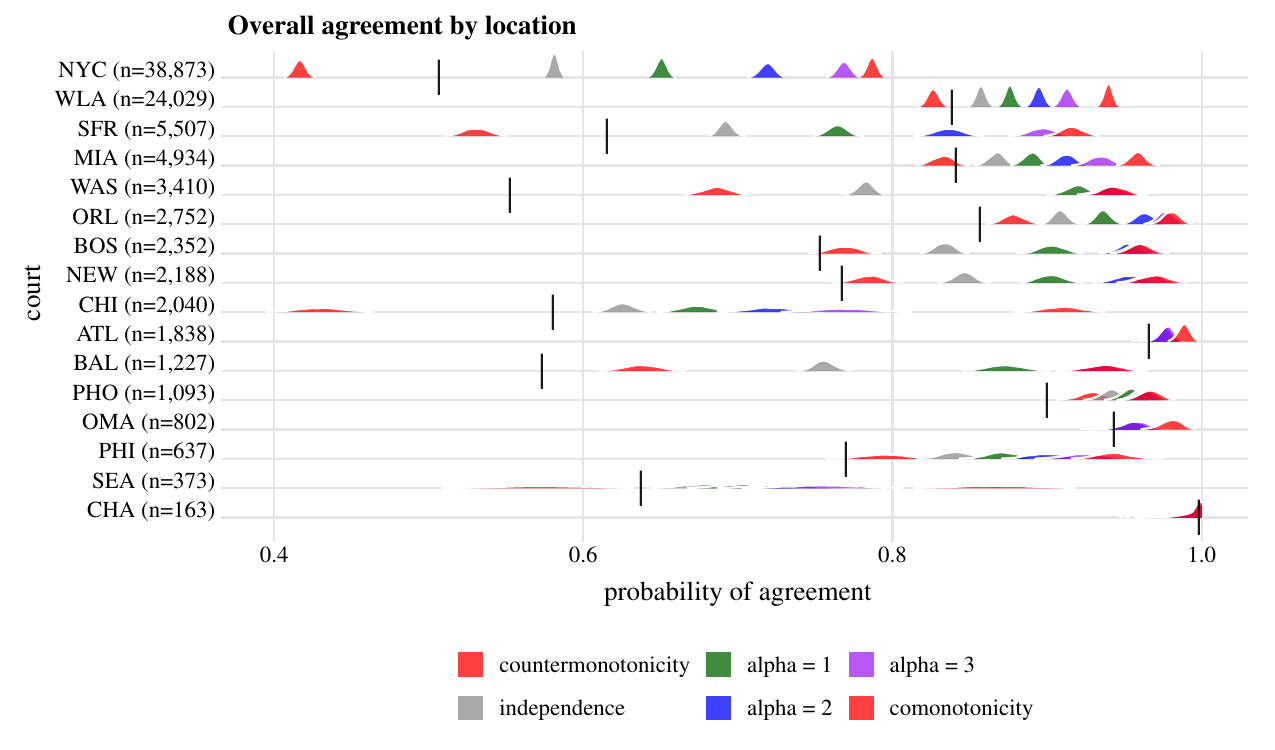}

    \end{subfigure}

    \begin{subfigure}{\textwidth}
        \centering
        \caption{Parametric}
        \label{fig:location_agreement_brms}
        \vspace{1em}
        \includegraphics[width=.65\linewidth,
                         trim= 0 0 0 .76cm, 
                         clip
                         ]{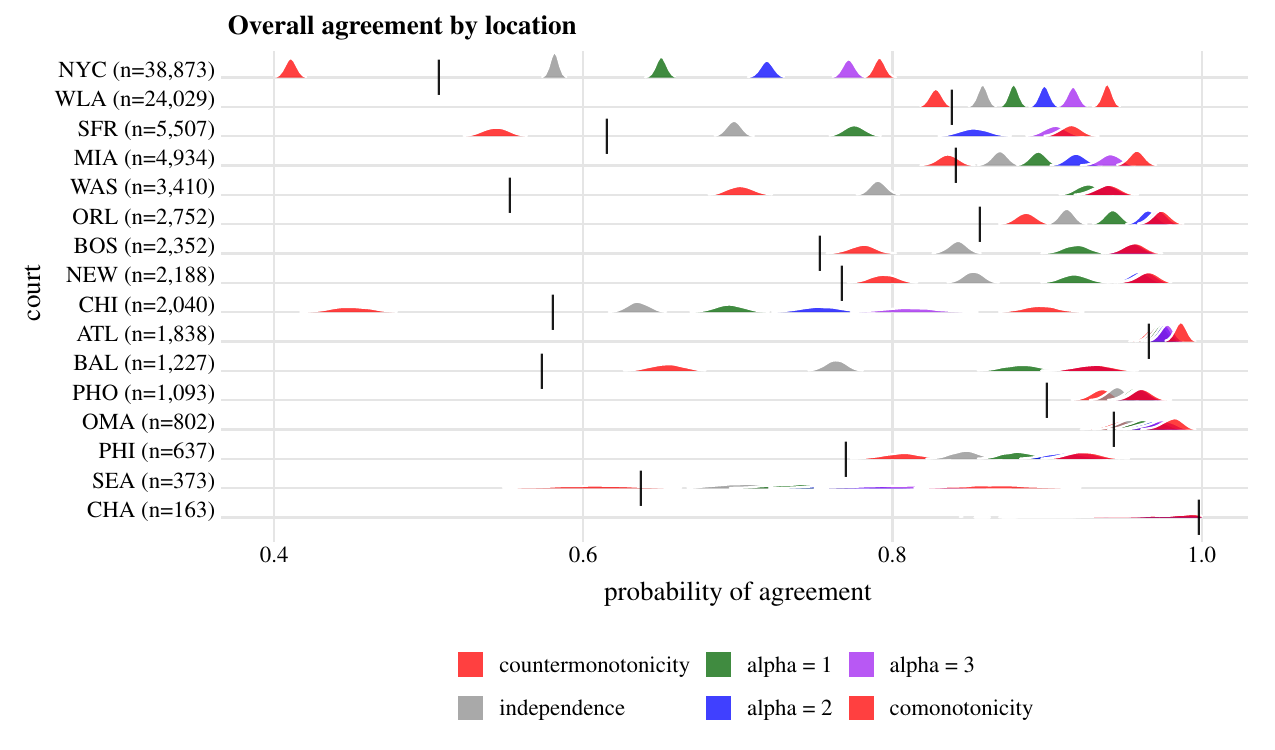}
    \end{subfigure}

\end{figure}    

\FloatBarrier

\printbibliography[title={E \quad References \vspace{.5em}}]
\end{refsection}

\end{document}